\documentclass[useAMS,usenatbib,usegraphicx]{mn2e}
\usepackage{times}
\usepackage{caption}

\def\swift{{\sl Swift}}

\def\integral{{\sl INTEGRAL}}
\def\astrosat{{\sl ASTROSAT}}

\def\fermi{{\em Fermi}}

\def\ultracam{{\sc ultracam}}
\def\p{$\pm$}
\def\ltsim{\mathrel{\hbox{\rlap{\hbox{\lower4pt\hbox{$\sim$}}}\hbox{$<$}}}}
\def\gtsim{\mathrel{\hbox{\rlap{\hbox{\lower4pt\hbox{$\sim$}}}\hbox{$>$}}}}

\def\Msun{M$_{\odot}$}

\def\aap{A\&A}
\def\nat{Nat}
\def\aaps{A\&AS}
\def\mnras{MNRAS}
\def\apj{ApJ}
\def\ssr{SSR}
\def\iaucirc{IAU\,Circ.}

\def\aj{AJ}
\def\apjl{ApJL}
\def\pasp{PASP}
\def\pasj{PASJ}
\def\procspie{SPIE}
\def\ha{H$\alpha$}
\def\hb{H$\beta$}

\def\gravrad{$R_{\rm G}$}

\def\mbh{$M_{\rm BH}$}
\def\gx339{GX\,339--4}
\def\v404{V404\,Cyg}
\def\swiftj1753{SWIFT\,J1753.5--0129}
\def\xtej1118{XTE\,J1118+480}
\def\av{$A_{\rm V}$}

\def\numax{$\nu_{\rm max}$}
\def\nub{$\nu_{\rm b}$}

\title[Multi-component optical variability in V404\,Cyg]{Furiously Fast and Red: Sub-second Optical Flaring in V404\,Cyg during the 2015 Outburst Peak}

\author[P. Gandhi et al.]{P. Gandhi,$^1$ 
S.P. Littlefair,$^2$ L.K. Hardy,$^2$ V.S. Dhillon,$^{2,3}$ T.R. Marsh$^4$, A.W. Shaw,$^1$\newauthor D. Altamirano,$^1$ M.D. Caballero-Garcia,$^{5}$  J. Casares,$^{3,6,7}$ P. Casella,$^8$\newauthor A.J. Castro-Tirado,$^{9,10}$ P.A. Charles,$^1$ Y. Dallilar,$^{11}$ S. Eikenberry,$^{11}$ R.P. Fender,$^7$\newauthor R.I. Hynes,$^{12}$ C. Knigge,$^1$ E. Kuulkers,$^{13}$ K. Mooley,$^7$ T. Mu\~{n}oz-Darias,$^{3,6}$ M. Pahari,$^{14}$\newauthor F. Rahoui,$^{15,16}$ D.M. Russell,$^{17}$ J.V. Hern\'{a}ndez Santisteban,$^1$ T. Shahbaz,$^{3,6}$\newauthor D.M. Terndrup,$^{18}$ J. Tomsick,$^{19}$ D.J. Walton$^{20}$\\
$^{1}$Department of Physics and Astronomy, University of Southampton, Highfield, Southampton SO17 1BJ\\
$^{2}$Department of Physics and Astronomy, University of Sheffield, Sheffield S3 7RH\\
$^{3}$Instituto de Astrofisica de Canarias, 38205 La Laguna, Santa Cruz de Tenerife, Spain\\
$^{4}$Department of Physics, University of Warwick, Gibbet Hill Road, Coventry, CV4 7AL\\
$^{5}$Astronomical Institute, Academy of Sciences of the Czech Republic, Bo$\check{c}$n\'{\i}~II 1401, CZ-141\,00~Prague, Czech Republic\\
$^{6}$Dept. Astrofísica Universidad de La Laguna (ULL), E-38206 La Laguna, Tenerife, Spain\\
$^{7}$Astrophysics, Department of Physics, University of Oxford, Keble Road, Oxford OX1 3RH, UK\\
$^{8}$INAF-Osservatorio Astronomico di Roma, Via Frascati 33, I-00040 Monteporzio Catone, Italy\\
$^{9}$Istituto de Astrofisica de Andalucia (CSIC), Granada E-18008, Spain\\
$^{10}$Unidad Asociada Departamento de Ingenieria de Sistemas y Automatica, Universidad de Malaga, Malaga, 29071, Spain\\
$^{11}$Department of Astronomy, University of Florida, 211 Bryant Space Science Center, Gainesville, FL 32611, USA\\
$^{12}$Department of Physics and Astronomy, Louisiana State University, 202 Nicholson Hall, Tower Drive, Baton Rouge, LA 70803, USA\\
$^{13}$European Space Astronomy Centre (ESA/ESAC), Science Operations Department, 28691 Villanueva de la Cañada, Madrid, Spain\\
$^{14}$Inter-University Center for Astronomy and Astrophysics, Post Bag 4, Ganeshkhind, Pune-411007, India\\
$^{15}$European Southern Observatory, K. Schwarzschild-Strasse 2, D-85748 Garching bei M\"{u}nchen, Germany\\
$^{16}$Department of Astronomy, Harvard University, 60 Garden street, Cambridge, MA 02138, USA\\
$^{17}$New York University Abu Dhabi, PO Box 129188, Abu Dhabi, UAE\\
$^{18}$Department of Astronomy, The Ohio State University, 140 West 18th Avenue, Columbus, OH 43210, USA\\
$^{19}$Space Sciences Laboratory, 7 Gauss Way, University of California, Berkeley, CA 94720-7450, USA\\
$^{20}$Jet Propulsion Laboratory, California Institute of Technology, 4800 Oak Grove Drive, Mail Stop 169-221, Pasadena, CA 91109, USA
}

\begin{document}

\voffset=-0.50in

\date{\today}
\date{Accepted Mar 04 2016. 
      Revised Mar 03 2016. 
      Reviewed Feb 22 2016. 
      Received Dec 22 2015.}

\pagerange{\pageref{firstpage}--\pageref{lastpage}} 
\pubyear{2016}

\maketitle
\label{firstpage}

\begin{abstract}
  We present observations of rapid (sub-second) optical flux variability in V404\,Cyg during its 2015 June outburst. Simultaneous three-band observations with the ULTRACAM fast imager on four nights show steep power spectra dominated by slow variations on $\sim$\,100--1000\,s timescales. Near the peak of the outburst on June\,26, a dramatic change occurs and additional, persistent sub-second optical flaring appears close in time to giant radio and X-ray flaring. The flares reach peak optical luminosities of $\sim$\,few\,$\times$\,10$^{36}$\,erg\,s$^{-1}$. Some are unresolved down to a time resolution of 24\,milliseconds. Whereas the fast flares are stronger in the red, the slow variations are bluer when brighter. The redder slopes, emitted power, and characteristic timescales of the fast flares can be explained as optically-thin synchrotron emission from a compact jet arising on size scales $\sim$140--500\,Gravitational radii (with a possible additional contribution by a thermal particle distribution). The origin of the slower variations is unclear. The optical continuum spectral slopes are strongly affected by dereddening uncertainties and contamination by strong \ha\ emission, but the variations of these slopes follow relatively stable loci as a function of flux. Cross-correlating the slow variations between the different bands shows asymmetries on all nights consistent with a small red skew (i.e., red lag). X-ray reprocessing and non-thermal emission could both contribute to these. These data reveal a complex mix of components over five decades in timescale during the outburst. 

\end{abstract}

\section{Introduction}

\v404\ is an X-ray binary (XRB) hosting a black hole (BH) with mass in the range \mbh\,=\,8--15\,\Msun\ \citep[e.g. ][ and references therein]{khargharia10, casares14}. It came to great prominence in 1989 as the X-ray nova GS\,2023+338 \citep{makino89}. Thereafter, it was found to be associated with prior optical nova eruptions of the source \v404\ in 1938 and 1956 \citep[e.g. ][]{richter89}, suggesting repeated long-term outbursts. In June 2015, the source underwent a brief but prolific outburst, during which it reached peak brightness levels amongst the brightest in the X-ray sky, with luminosities rivalling the Eddington value during flares \citep{ferrigno15, rodriguez15}. Its brightness makes this outburst an excellent laboratory for studying massive accretion episodes on to black hole XRBs in detail. 

Multiwavelength variability can provide important clues to the physical origin of emission in XRB outbursts. Several studies in recent years have pointed out that rapid multiwavelength timing observations can break degeneracies between emission models comprising the accretion disc, the inner hot flow, and the base of the jet \citep[e.g. ][ and references therein]{kanbach01, uemura02_v4641sgr, hynes03_xtej1118, durant08, g10, casella10, veledina11, malzac14, drappeau15, uttley14_review}. Yet, the number of black hole XRBs with detections of sub-second variability remains just a few.

The June\,2015 outburst of \v404\ was also accompanied by spectacular multiwavelength variability across the entire electromagnetic spectrum. The fastest characteristic variability timescales reported so far at any wavelength other than X-rays go down to $\sim$1\,s \citep[][ both in the optical]{hynes15_atel2, terndrup15}. Although faster observations have been carried out, there has been no report of the detection of significant sub-second flaring so far \citep[cf., ][]{g15_atel1}. 

In this paper, we present rapid, sub-second imaging observations of \v404\ carried out with the ULTRACAM fast imager \citep{ultracam} mounted on the 4.2\,m William Herschel Telescope (WHT) in La Palma. The data were obtained over four nights and lead up to the night of 2015\,June\,25 (2015\,June\,26\,UT) when the source showed particularly intense radio and X-ray flaring activity. We find persistent sub-second optical flaring activity on this night, and contrast this with the preceding nights which were dominated mainly by slower variations with characteristic timescales of several hundred seconds and longer. This paper places some first constraints on the optical emission processes of the fast flares using timing and spectral analyses, together with qualitative discussion of the slow variations. More detailed investigations will be presented in follow-up works. 

\section{Observations}

\subsection{Fast optical timing with ULTRACAM/WHT}
The observations were carried out over several nights during the week of 2015 June 19--26, for durations ranging from $\sim$20\,min to 1.5\,hr. Observations on four nights were obtained under clear skies at low airmass ($\le$\,1.1), with photometry being carried out simultaneously in the $u'$, $g'$ and $r'$ bands. The observation log for these nights is presented in Table\,\ref{tab:log}. Other than one night that was pre-scheduled and coordinated with X-ray observations with the \swift\ satellite, all observations were carried out on a best-effort basis during gaps in the normal ULTRACAM observing programme, sometimes extending into morning twilight in order to maximise the duration of the observing window. A few additional short observations were carried out either under worse weather conditions or in non-standard filters, and are not analysed herein. 

ULTRACAM was used in its fastest \lq drift\rq\ mode, except on the first night when \lq one-pair\rq\ mode was employed. These modes allow fast, simultaneous photometry within small CCD windows centred on the target and a field comparison star within a few arcmin of V404\,Cyg. The window sizes were typically $\sim$\,50\,$\times$\,50\,pixels. Cycle times depend upon window size and CCD binning, and range from 78\,ms down to 24.1\,ms. 
Dead time was small for most of the observations which were carried out in drift mode ($\approx$\,1.1\,ms), and is ignored in our analysis here. For the first observation in one-pair mode, dead time is higher ($\approx$\,24\,ms) and constitutes 30\,\% of the cycle time. Its impact is noted where relevant (specifically for the power spectra that we will present later). 

The field star was used to monitor seeing and transparency variations. But \v404\ in outburst was brighter than all stars in the ULTRACAM field of view, especially in the redder $r'$ and $g'$ filters where we have the fastest sampling, so the field star is not used for relative photometry. 
The $u'$ band is less sensitive than the other two, and coadding of frames is employed to provide a gain in signal-to-noise (S/N). The coadding is performed on-chip, and individual coadded frames are not saved. On the first night, 6 frames were coadded, and the other nights employed 15 coadds in $u'$, resulting in correspondingly lower time resolution.

Data reduction was carried out with the ULTRACAM pipeline v.9.14 \citep{ultracam}. All frames were bias-subtracted and flat-fielded. Source photometry was carried out in large circular apertures (7--14\,arcsec diameter) with variable centre positions tracking the centroid of the source on each frame. 
The seeing was typically between 1--2\,arcsec during the observations. 
Sky background was measured as the clipped mean in an annular aperture. 

Photometric errors include Poisson noise and read noise. Photometric calibration was not the focus of these observations, whose purpose was to search for optical variability. However, the nights were clear and an approximate flux calibration was possible using zeropoints based upon photometric standard stars measured on some of the nights. The scatter between the nights is $\approx$\,0.04, 0.005 and 0.01\,mag, respectively, in $u'$, $g'$ and $r'$. Photometric calibration systematic uncertainties and tests are described in the Appendix. However, note that it is the systematic uncertainties related to the large line-of-sight reddening that dominate measurements of the intrinsic shape of the optical spectral energy distribution (SED), as we will discuss later. 

\subsection{Optical spectroscopy with BOOTES}

Broadband filters are used in ULTRACAM, with wavelength coverage including contributions from emission lines in addition to the spectral continuum. The most prominent optical emission line in the case of V404\,Cyg is \ha, which falls in the $r'$ band. In order to estimate the relative flux contribution of \ha, we utilised optical spectroscopic data from BOOTES--2/COLORES. As we will discuss later, the emission line strength relative to continuum is known to be strongly variable, so only (quasi)simultaneous data are appropriate for such estimates. 

BOOTES (acronym of the {Burst Observer and Optical Transient Exploring System}) is a world-wide network of robotic telescopes \citep{bootes,castrotirado12}, with telescopes located in Spain ({BOOTES}--1, {BOOTES}--2 and {BOOTES}--IR), New Zealand ({BOOTES}--3) and China ({BOOTES}--4). Currently, one optical spectrograph is operational in the {BOOTES} network. This is COLORES, mounted on the 0.6\,m diameter {BOOTES}--2 telescope. {COLORES} stands for {Compact Low Resolution Spectrograph} \citep{rabaza14}. It is sensitive over the wavelength range of 3800--11500\,\AA\ and has a spectral resolution of 15--60\,\AA. The primary scientific target of the spectrograph is prompt follow-up of Gamma Ray Bursts, but it is also used to study optical transients.

BOOTES--2/COLORES observed V404\,Cyg on several nights during the 2015 outburst for 300\,s of integration each time \citep{caballerogarcia15}. Data were reduced using standard procedures and wavelength-calibrated using arc lamps. The highest S/N spectrum was obtained on June\,26, starting at UT03:45:13. This is just over 1\,h before the ULTRACAM observations on this night (Table\,\ref{tab:log}). It is important to note that the spectra are not flux calibrated, so estimation of the relative strength of \ha\ can only be made by modelling the continuum.

\begin{table*}
\begin{center}
  \begin{tabular}{lcccccr}
    \hline
    Start Time & MJD & Duration & $g'r'$ Cycle time & $g'r'$ Exposure time & \# of $g'r'$ frames & $u'$ Coadd factor \\
       UTC     &  UTC &     s    &     ms    &     ms  &    &    \\
       (1)     &  (2) & (3) & (4)       & (5)     & (6)& (7)\\
    \hline
    June 20\,\,\,\,\,\,\,\,\,\,\,\,\,\,\,\,\,\,\,\,\,\,04:10:34.1 & 57193.17400622    &  4442.1    & 77.8 & 53.8 & 57065 & 6\\
    June 21\,\,\,\,\,\,\,\,\,\,\,\,\,\,\,\,\,\,\,\,\,\,03:40:23.6 & 57194.15305147    &  5900.0    & 35.9 & 34.7 & 164500 & 15\\
    June 25\,\,\,\,\,\,\,\,\,\,\,\,\,\,\,\,\,\,\,\,\,\,03:37:08.7 & 57198.15079469    &  6481.4    & 35.9 & 34.8 & 180350 & 15\\
    June 26\,epoch\,1\,\,\,04:52:35.4 & 57199.20318799    &  385.2     & 35.9 & 34.8 & 10718 & 15\\
    June 26\,epoch\,2\,\,\,05:00:08.8 & 57199.20843491    &   788.6    & 24.1 & 22.9 & 32780 & 15\\
    \hline
  \end{tabular}
  \caption{ULTRACAM Observations log\label{tab:log}. Columns\,(1) and (2) show the start times of the observations, and Col.\,(3) the duration of the light curves. The effective cycle time between consecutive frames and the integration (exposure) time of each frame are in Cols.\,(4) and (5), respectively. The dead time is the difference between these, and is large only for June 20. The number of frames for $g'r'$ in each observation is tabulated in Col.\,(6). The $u'$ band has a lower number of frames because of the on-chip coadding factor, listed in Col.\,(7).}
\end{center}
\end{table*}

\section{Results}

\subsection{Light curves from all nights}
\label{sec:lc}

\begin{figure*}
  \begin{center}
    \includegraphics[height=23cm,angle=-180]{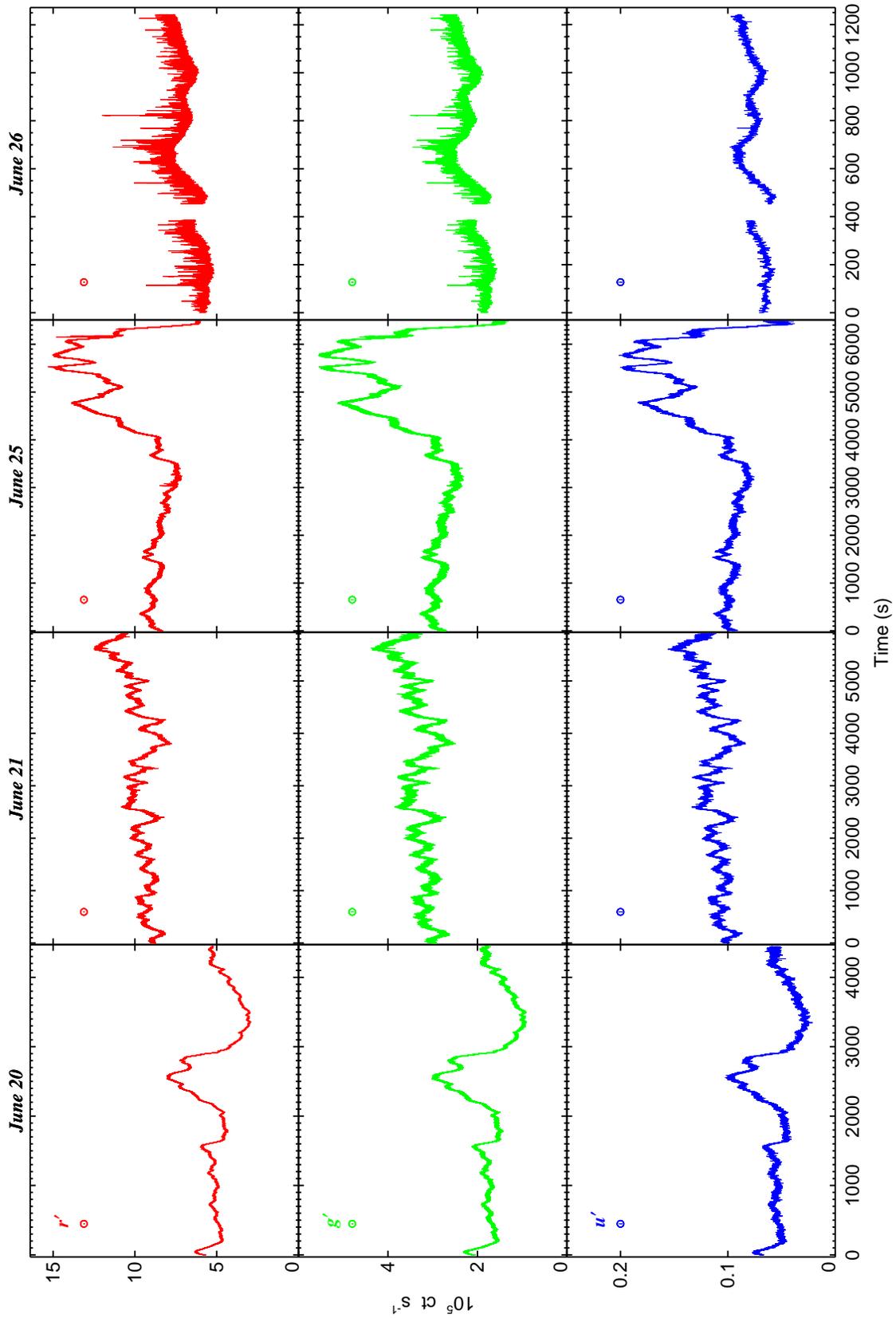}
    \caption{ULTRACAM light curves in ct rate units for all nights. Time on the x-axis is relative to the start of the observation on that night. The main features that stand out are smooth and \lq slow variations\rq\ on all nights, and \lq fast sub-second flares\rq\ mainly on the last night on June\,26. The second night of June\,21 additionally shows an oscillatory pattern on intermediate timescales of a few hundred seconds. The median size of the statistical errors for all light curves is denoted by the small vertical bar within the circles below the filter names. 
 \label{fig:lcall}}
  \end{center}
\end{figure*}

The source light curves are shown in Fig.\,\ref{fig:lcall}. \v404\ is significantly detected in all bands and on all nights, with count rates varying by about a factor of 5 in each band across all observations. The source count rates were found to be high in all cases, exceeding 10$^3$\,cts\,s$^{-1}$ in $u'$ and 10$^5$\,cts\,s$^{-1}$ in the other two bands. The S/N of the source is above $\approx$\,100 per frame in all $r'$ light curves. The S/N in $g'$ and $u'$ dips as low as $\approx$\,10 during some of the faintest short periods, but is typically higher by a factor of a few, at least. 

Strong and smooth variations on characteristic timescales of $\sim$\,tens of minutes are present on all nights. We generically refer to these as the \lq slow variations\rq\ hereafter. Some qualitative changes are visible between the first three nights, with the data on June\,21 (here and hereafter, the night of observation is referred to in UTC) showing a quasi-regular oscillatory pattern on timescales of a few minutes which are absent on the other nights. The most striking change, however, occurs on the last night of June\,26. In addition to the slow variations, the light curve on this night is crowded with short, spiky flares (hereafter, \lq fast flares\rq\ or \lq sub-second flares\rq) for the entire duration of about 1200\,s. By contrast, the first two nights show no such sharp flares, while the third night (June\,25) shows only a few, isolated periods of flaring, discussed later. 

\begin{figure*}
  \begin{center}
    \includegraphics[angle=90,width=18cm]{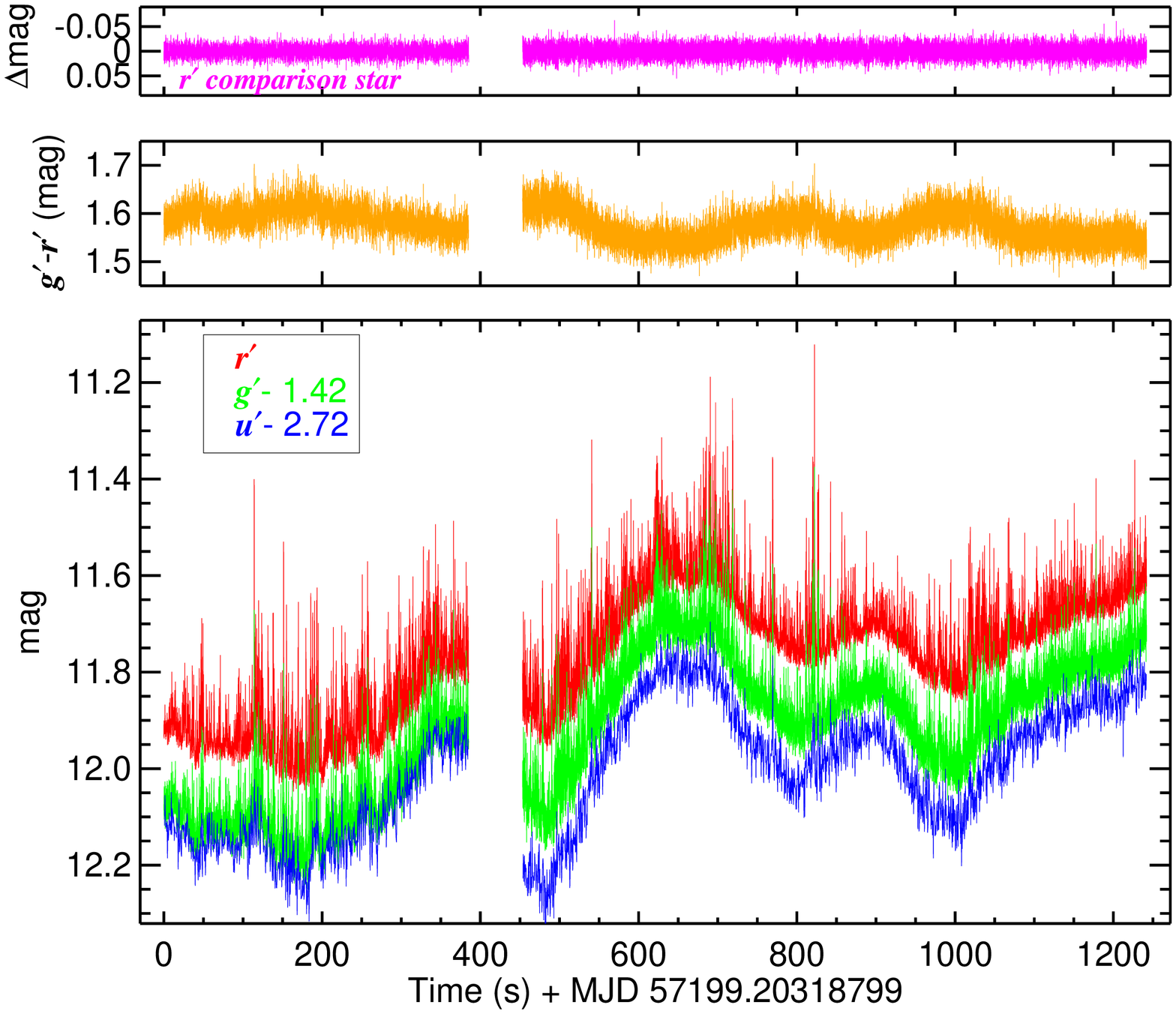}
\caption{ULTRACAM light curves from 2015 June\,26 (UTC) of V404\,Cyg (main bottom panel). The first (0--385\,s) and the second (453--1242\,s) epochs of observation were separated by a short gap. The light curves have been offset slightly for clarity. This figure clearly shows the high density of fast flaring activity, which is strongest in $r'$, followed by $g'$.  
The Top panel shows the comparison star light curve (only $r'$ is shown for clarity); its small scatter (1.11\% around mean) demonstrates the stability of observing conditions. The middle panel shows the source colour for the $g'$ and $r'$ bands which have the highest time resolution. We remind the reader that the $u'$ light curve has lower time resolution by a factor of 15 than the other two bands, so is not suitable for comparing the fast flares with the other bands. Comparing the middle and bottom panels, the slow variations are seen to be bluer (smaller $g'$--$r'$ colour) when the source is brighter, and vice-versa. The instants of the fast flaring are also associated with momentary reddening. 
 \label{fig:lc}}
  \end{center}
\end{figure*}

Fig.\,\ref{fig:lc} enlarges the flux-calibrated June\,26 lightcurves from all three bands for clearer comparison. 
After an initial observation 385.2\,s in length (hereafter, epoch 1) on this night showed the presence of obvious fast flaring activity, the observation was paused, sped up, and continued for a further duration of 788.6\,s (epoch 2). The $g'$--$r'$ colour evolution is also plotted. The slow variations are seen to have stronger peak-to-peak variability in the bluer filters. This is apparent from the fact that the colour is bluer (smaller $g'$--$r'$) when the source is brighter, and redder when fainter. In fact, this is true for all nights of data, as we will discuss later.\footnote{Also apparent in the overlays presented by \citet{g15_atel1}.}

In contrast to the stronger blue variability of the slow variations, the figure clearly shows that the fast flaring is stronger in $r'$ than in $g'$. This is apparent from strength of the $r'$ flares in the main panel, and the momentary reddening ($g'$--$r'$ colour spikes) at the times of the fast flares in the middle panel. We remind the reader that the $u'$ band has an intrinsic time resolution 15\,times lower than the other bands on this night, so is unsuitable for comparisons on the fastest timescales. We will analyse the properties of the fast flares and colour trends in more detail shortly. We also refer the reader to the supplementary Appendix Fig.\,\ref{fig:lcoverlay} which illustrates this further.

It is worth noting that the sub-second flaring is clearly much stronger than atmospheric scintillation noise, whose magnitude can be gauged from the simultaneously-observed comparison star light curve in the same figure. The comparison star shows a fractional r.m.s. ($F_{\rm var}$; \citealt{vaughan03}) of 0.010--0.016 across the bands. This r.m.s. will include contributions from both Poisson and scintillation noise. $F_{\rm var}$ for the source is much stronger, at 0.075--0.096, over the entire light curve, i.e. over the full Fourier frequency range (see following sub-section).  

\subsection{Power spectra}
\label{sec:psd}

\begin{figure*}
  \begin{center}
    \includegraphics[angle=0,height=6cm]{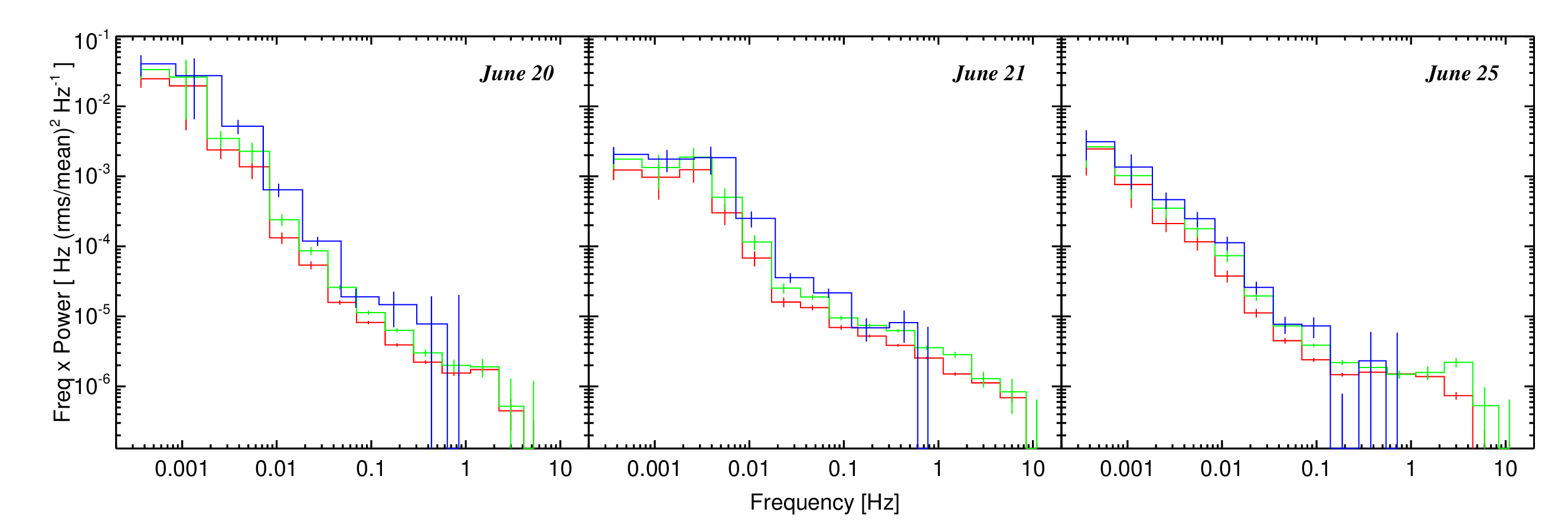}
    \includegraphics[angle=0,height=6cm]{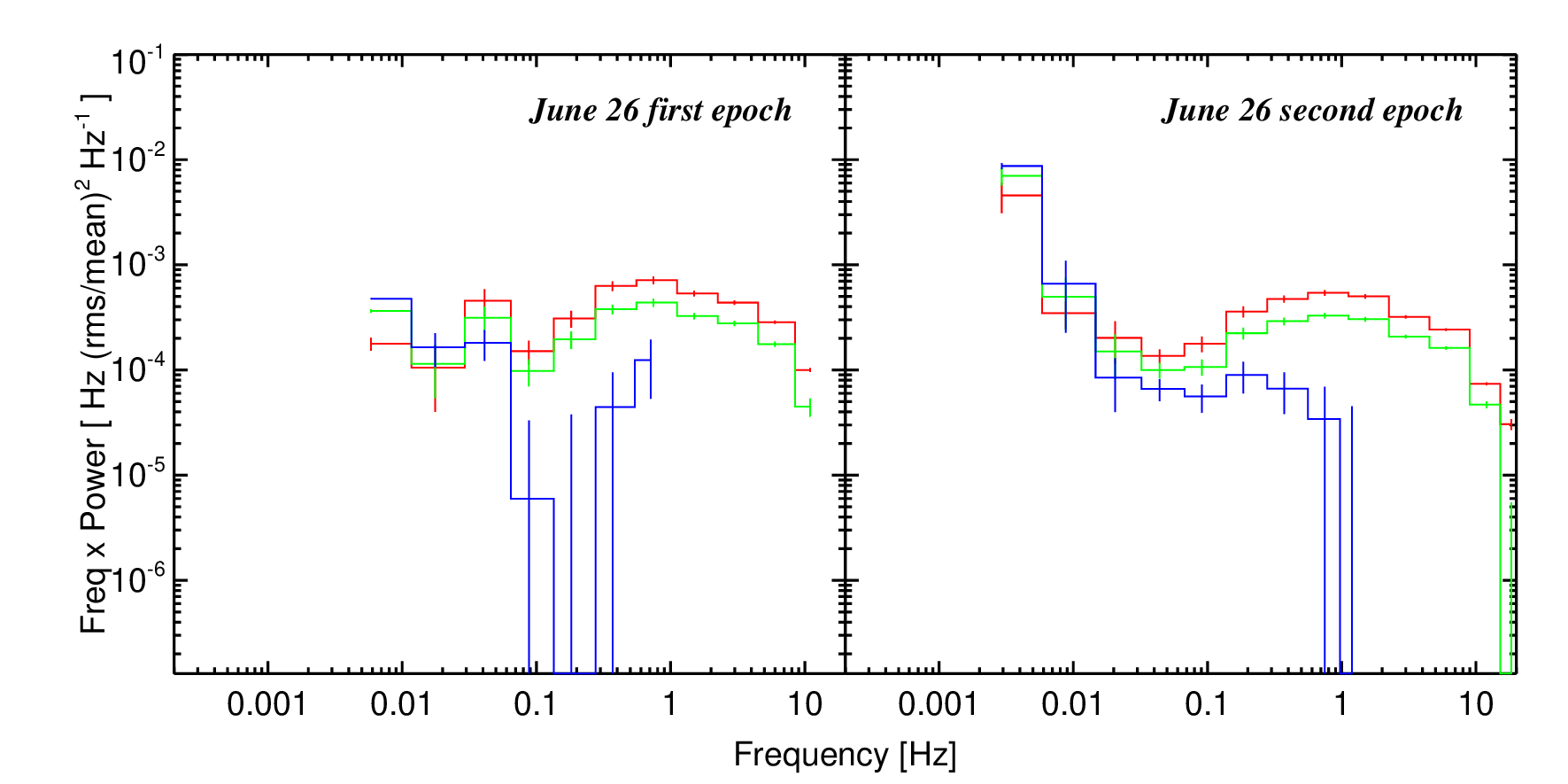}
\caption{$u'g'r'$ power spectra (PSDs) in $\nu P_\nu$ units for all nights compared on an identical scale. Colours are as in Fig.\,1. The slow (low Fourier frequency) variations are stronger in the bluer bands on all nights. The high frequency (fast flaring) behaviour is dramatically different in both strength and colour on June\,26. 
 \label{fig:powspec_all}}
  \end{center}
\end{figure*}

Fig.\,\ref{fig:powspec_all} shows the power spectral densities (PSDs) in rms-normalised $\nu P_\nu$ (Frequency\,$\times$\,Power) units. The PSDs for the observations on all nights are compared on an identical scale. PSDs were computed over long sections of 4096\,s for the first three nights, and shorter sections of 256\,s and 512\,s respectively for epochs 1 and 2 on the last night, according to the length of the observations. 
The PSDs were corrected for constant white noise at high Fourier frequencies using standard formalisms \citep{vaughan03}. In addition to photometric measurement noise, scintillation noise becomes non-negligible at high Fourier frequencies for ground-based observations. \citet{osborn15} provide estimates of the magnitude of scintillation noise for the La Palma site, and we used their Eq.\,7 to add in this noise component in quadrature for the computation for white noise. However, the estimates of \citeauthor{osborn15} are based upon median atmospheric turbulence profiles, so may overestimate the true white noise component in case of low turbulence. For cases where the estimated white noise level exceeded the measured PSD at high Fourier frequencies, we instead used a constant to model the PSD level at the three highest frequencies and used this as a measure of the white noise. Finally, a logarithmic frequency binning was applied. 

The figure shows steeply falling PSDs over the first three nights. Fitting a simple power law model $P_{\nu}$\,$\propto$\,$\nu^{\beta}$ to these PSDs, 
we find typically slopes $\beta$ ranging from $\approx$\,--1.6 to --2 below $\sim$\,1\,Hz. At the lowest Fourier frequencies $\ltsim$\,0.01\,Hz, the PSD bends into an even steeper component. 
This second component appears especially prominent on the second night of June\,21, when it levels off on timescales of $\sim$\,500\,s 
and is likely associated with the repeated oscillatory behaviour visible in the lightcurve in Fig.\,\ref{fig:lcall}. 
In all these cases, power increases by a small factor in the bluer bands (typically 1.3 in $g'$ and 1.7 in $u'$, respectively, relative to $r'$). 
These are again a reflection of the stronger peak-to-peak slow variability in the bluer bands apparent in the light curves and highlighted in the previous section. The effect of dead time is to increase the noise floor and alias power towards low Fourier frequencies because some of the high frequency power is not directly sampled. This would affect mainly the first night of data (obtained in \lq one-pair mode\rq; see Observations section) when dead time is significant. Despite this effect, the behaviour of the PSDs on this night appears very similar to the other nights. Moreover, a steep red noise optical PSD has also been reported during the present V404\,Cyg outburst by \citet{hynes15_atel1} and \citet{wiersema15}. Detailed investigations of these PSDs sampling the slow variations will be presented in later work. 

Here, we focus on the PSDs from the last night of June\,26 which shows fast flaring activity. The PSDs in the bottom panels of Fig.\,\ref{fig:powspec_all} show a striking difference with respect to the previous PSDs, in that they now display a strong and broad hump at the highest Fourier frequencies, extending over $\sim$\,0.1--10\,Hz. This hump is completely absent on the other nights, and the peak power in the hump is about 100--200\,times higher than the power seen during the previous nights at $\sim$\,1\,Hz. Furthermore, it is now the redder bands that carry more power above $\sim$0.05\,Hz. At lower frequencies, there is a changeover to the bluer bands dominating the slow variations -- a behaviour that is similar to the previous nights. This changeover is best visible for the second epoch, where the light curve duration is longer. This also has finer sampling, allowing the PSD to be generated over a broad range from 0.002\,Hz up to the Nyquist frequency of $\approx$\,20\,Hz. A steep up-turn is apparent below the hump around a frequency of 0.01\,Hz. We found that when fitting the hump with a Lorentzian component, the lower frequency part of the PSD required two power laws to describe. In Table\,\ref{tab:psdfit}, we show the parameters for a double power law plus zeroth order Lorentzian fit to the $g'$ and $r'$ data for this epoch. One of the power laws is found to be very steep. We caution, however, that the relatively short duration of our light curve does not allow us to model the low frequencies well, and our model should only be taken as a parametrization of the PSD shape. The $u'$ light curve was slower and the PSD has lower S/N, but it qualitatively obeys the same trend as $g'$ and $r'$. 

By integrating the rms-normalised PSD over any given frequency range, the fractional variance of the light curve over that range can be determined. For the sub-second variability which is the focus of our work here, the fractional r.m.s. over 1--20\,Hz for the fastest June\,26 epoch 2 light curve constitutes 20\,\% and 12\,\% of the total fractional r.m.s. in $r'$ and $g'$ reported at the end of \S\,\ref{sec:lc}, respectively. 

\begin{table*}
  \begin{center}
 \begin{tabular}{lccccr}
    \hline
Band     & $\beta_1$ & $\beta_2$ &      $r$     &    \numax   &       $\chi^2$/dof \\
         &            &            &              &         (Hz)    &                    \\
    \hline                  			         			     
  $r'$   &  --2.50$_{-1.61}^{+0.50}$     &  --0.77$_{-0.16}^{+0.90}$     &    0.06$\pm$0.01     &         0.70$_{-0.08}^{+0.09}$     &                    8.1/5\\
  $g'$   &  --2.89$_{-0.65}^{+0.19}$     &  --0.69$_{-0.21}^{+0.38}$     &    0.04$\pm$0.01     &         0.69$_{-0.08}^{+0.09}$     &                    5.9/5\\
        \hline
  \end{tabular}
  \caption{Double Power law + zero centred Lorentzian fits to the $r'$ and $g'$ second epoch PSDs for June\,26. The Lorentzian functional form is $P(\nu)=r^2\Delta/\pi/[\Delta^2+(\nu-\nu_0)^2]$, as defined by \citet{belloni02}. $\nu_{\rm max}=\sqrt{\nu_0^2+\Delta^2}$, where $r$, $\Delta$, and \numax\ refer to the integrated rms (over the full range of --$\infty$ to +$\infty$), half-width-at-half-maximum, and characteristic frequency, respectively. The slopes of the two power laws are $\beta_1$ and $\beta_2$. Errors are for a $\Delta$\,$\chi^2$\,=\,+2.71, or 90\% confidence for a single parameter of interest. For a zero centred Lorentzian, $\nu_0$\,=\,0, so $\Delta$\,=\,\numax. 
\label{tab:psdfit}}
\end{center}
\end{table*}

\subsection{Fast flares}
\label{sec:fastflares}

\begin{figure*}
  \begin{center}
    \includegraphics[angle=90,width=18cm]{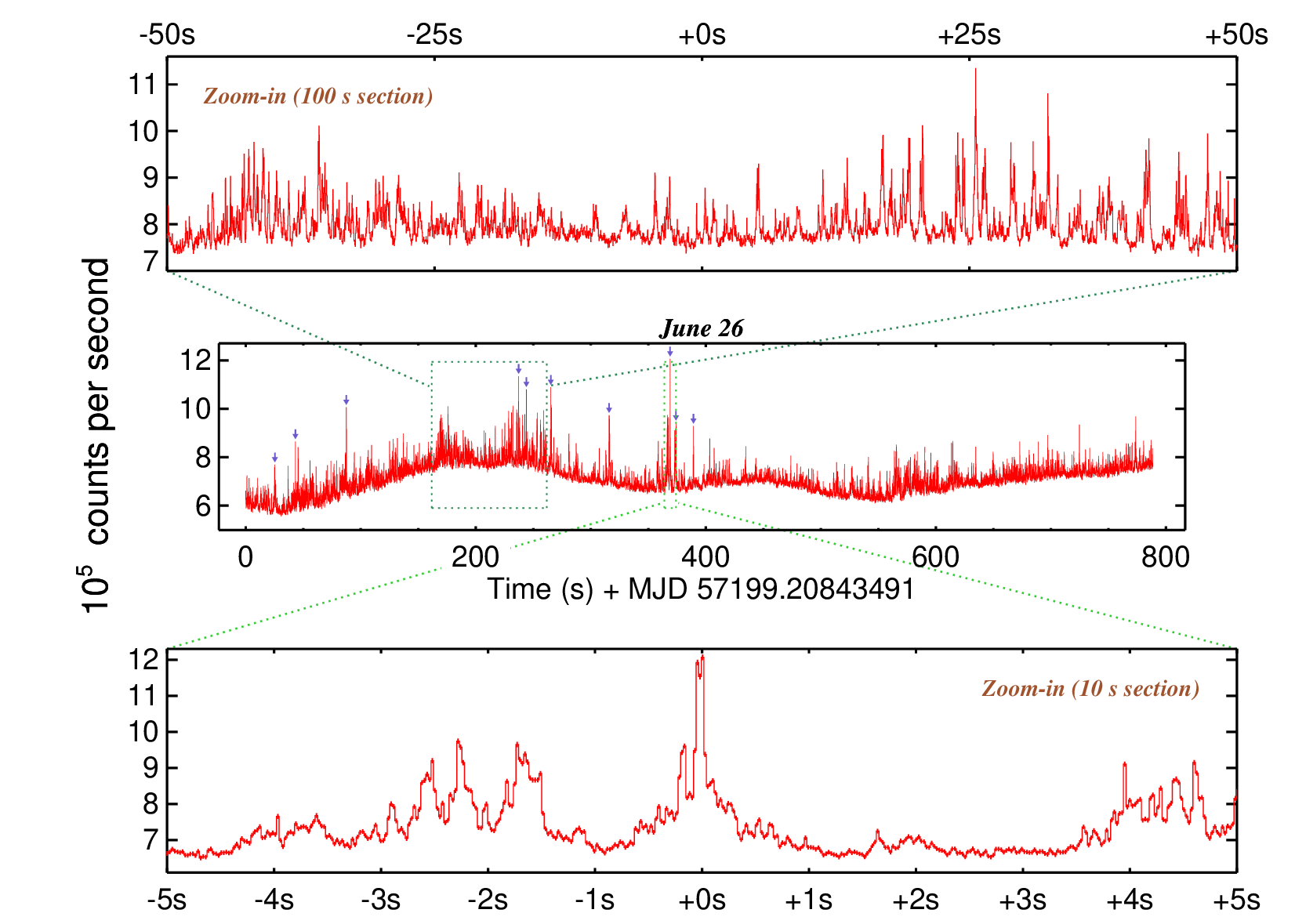}
\caption{Focusing on the fast red flares, this figure shows the full epoch 2 $r'$ light curve (central panel), together with zoom-ins of sections 100\,s (Top) and 10\,s (Bottom) in length. Statistical flux errors are less than 1\,\% and are plotted on the data points in the bottom panel. Scintillation noise is of the same order. The times of the 10 brightest flares selected in \S\,\ref{sec:fastflares} are marked by the violet arrows in the central panel. The lower zoom-in panel includes the very brightest of these flares. The native time resolution is $\approx$\,24\,ms in all panels. 
 \label{fig:lc2}}
  \end{center}
\end{figure*}

We next investigated the properties of the fast flaring behaviour on June\,26 in more detail. The high density of flaring makes it difficult to see the flares clearly in Fig.\,\ref{fig:lc}. So in Fig.\,\ref{fig:lc2}, we present zoom-in plots for the epoch 2 $r'$ light curve, concentrating on this band because it is the light curve with the strongest and fastest flaring.

\begin{figure}
  \begin{center}
    \includegraphics[angle=0,width=8.5cm]{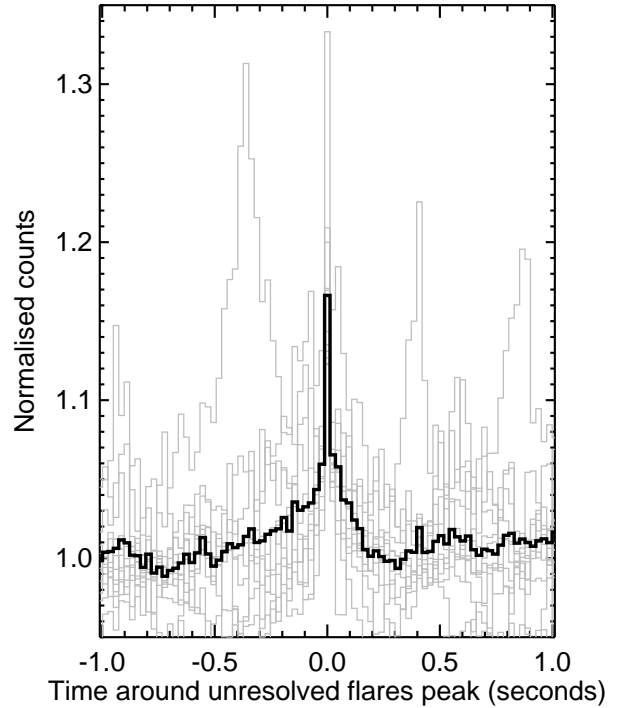}
    \caption{
      Twelve flares dominated by their unresolved narrow cores shown in grey, and their mean in bold black.  
 \label{fig:unresolvedflares}}
  \end{center}
\end{figure}

\begin{figure*}
  \begin{center}
    \includegraphics[angle=90,width=8.5cm]{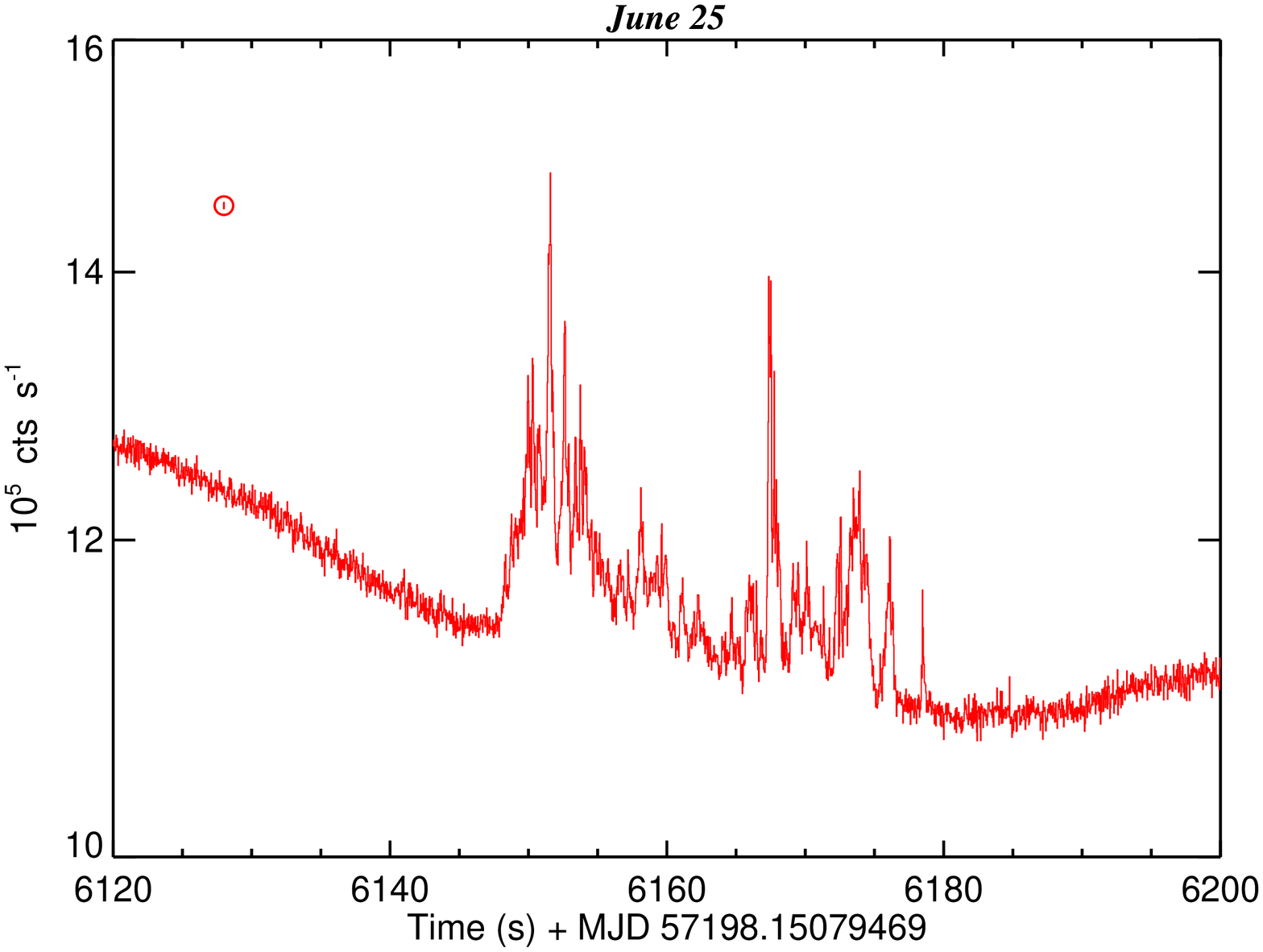}
    \includegraphics[angle=90,width=8.5cm]{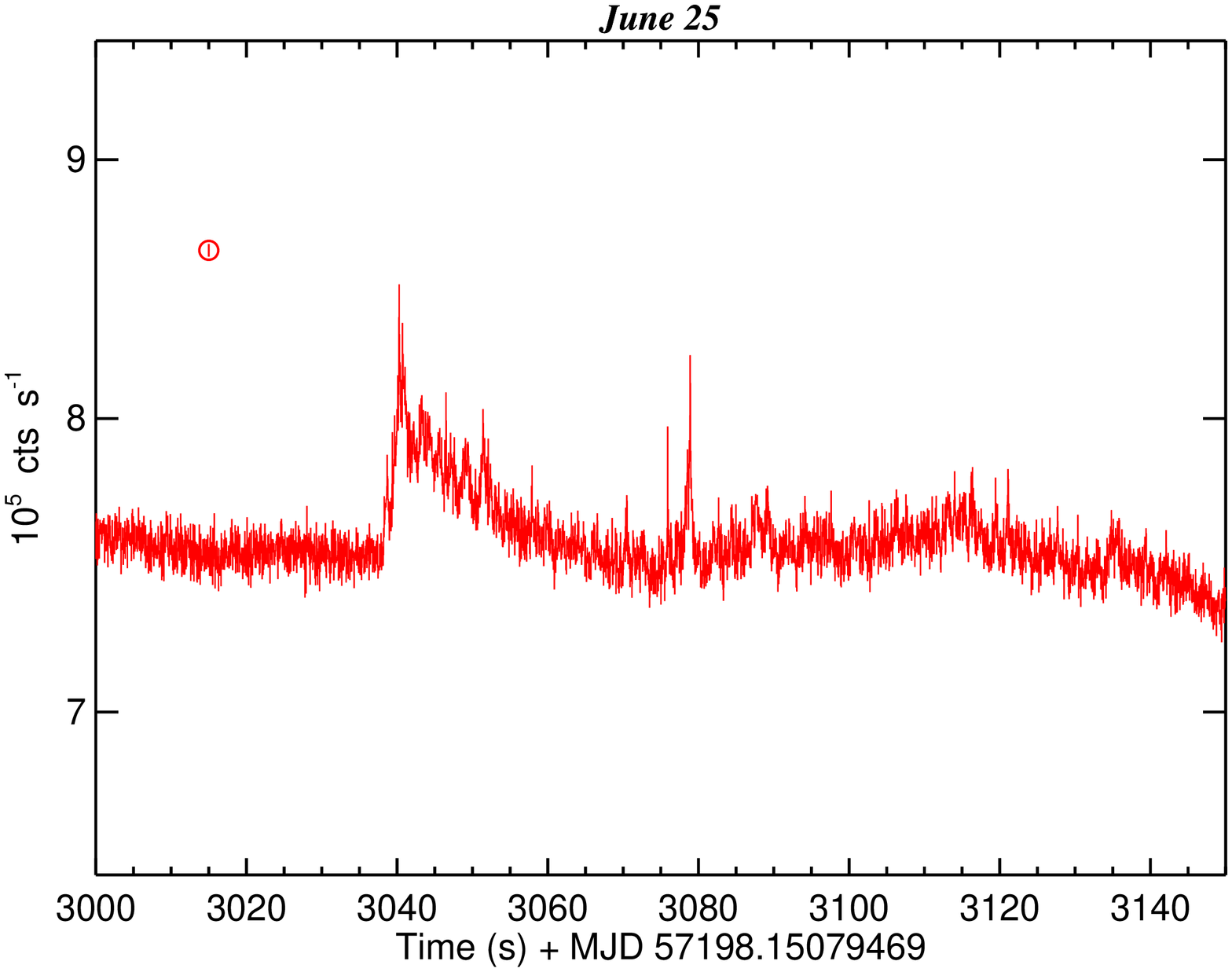}
\caption{Zoom-in plots around the two sections of the ULTRACAM $r'$ light curve from June\,25 showing short-lived sub-second flaring activity. Time is relative to the start of the observation on June\,25 (see Fig.\,\ref{fig:lcall}). Median statistical errors are plotted within the small circles.  
 \label{fig:lc3}}
  \end{center}
\end{figure*}

The PSDs for the same data set (Fig.\,\ref{fig:powspec_all}) show that flaring on a range of short timescales $\ltsim$\,10\,s manifests as a broad continuum hump in the frequency domain above $\sim$\,0.1\,Hz. Here, we examine the properties of these fast flares in the time domain. We started by detrending the light curve in order to normalise out the slow variations. This was done by using a running 20\,s clipped mean to define the local continuum count rate at every time bin, and then dividing by this continuum count rate to create a normalised light curve. A flare was then defined as one showing at least a 10\,$\sigma$ fluctuation in the normalised count rate light curve, where $\sigma$ here refers to the background scatter introduced by atmospheric scintillation, measured in the comparison star light curve. A selected flare was also required to be the brightest point within a contiguous interval of 10\,s ($\Delta t$\,=\,\p\,5\,s around the peak). This was done in order to select unique flares only and avoid double-counting in case of complex flare profiles. The choice of 10\,s as the interval was motivated by the lower frequency threshold of $\sim$\,0.1\,Hz above which the Lorentzian hump dominates the PSDs. 

We found 74 flares using the above criteria, with a mean flare peak strength of 23\,\% above the continuum count rate, and with an average time interval between flares of 10.6\,s. The brightest flare occurred 369.97\,s after start and showed a peak strength $\approx$\,78\%\ in count rate (or 0.63\, in mag). The lower zoomed panel in Fig.\,\ref{fig:lc2} shows this flare in close detail, together with other nearby features. At half maximum height, the full width of this flare is $\approx$\,70\,ms. 
Most flares, including this one, show complex profiles with a broad base and/or multiple neighbouring sub-flares around a narrow core. 
We select the 10 brightest flares (highlighted in the central panel of Fig.\,\ref{fig:lc2}) for characterisation. Fitting a Gaussian to the mean profile, we find a narrow core with a full width at half maximum (FWHM) 
of 345\,(\p\,1)\,ms. But as the PSD shows, there is significant flaring on timescales even shorter than this. For instance, choosing $\Delta t$\,=\,1 s (i.e. a contiguous interval of 2\,s for unique flare selection), we find 172 flares with a time interval between flares of 4.6\,s. Given the complex flare profiles, even higher flare rates can be inferred by using other methods to define and count flares. So the brightest selected flares do not necessarily tell us about the fastest flaring activity. 

We investigated this further by selecting flares that are dominated by an {\em unresolved} narrow core. 
This was done by requiring that the central single 24\,ms time bin at flare peak be at least a factor of 2 higher (above the continuum count rate) than its immediately adjacent bins on either side, in addition to being the brightest point within \p\,1\,s. 
Twelve such flares were found, and are plotted aligned to their peaks and superposed in Fig.\,\ref{fig:unresolvedflares}. The profiles are again complex, with sub-flares visible even within \p\,1\,s. But the mean profile clearly shows an unresolved peak dominating above a broad base. This mean profile peaks 17\,\% above continuum. So there are at least some flares with cores apparently narrower than 24\,ms. 

Finally, we highlight two short periods of significant sub-second flaring which occurred on the preceding night of June\,25. These are presented in Fig.\,\ref{fig:lc3}. These flaring episodes are concentrated within $\sim$\,30--40\,s and are remarkable in their sudden appearance and disappearance in the otherwise smooth light curve dominated by the slow variations. We examined the relative strengths of these flares between the bands, and found that they show stronger flaring in $r'$ than in $g'$, very similar to the fast persistent flaring on the final nights. No such episodes were detected during the observations on our first two nights (June\,20 and 21).

\subsection{Spectral indices of the fast flares and the slow variations}
\label{sec:colours}

In sections\,\ref{sec:lc} and \ref{sec:psd}, we found that the fast flares are stronger in the red than the slower variations. An examination by eye of the strongest flares in Fig.\,\ref{fig:lc} shows that they momentarily redden the source colour by up to $\Delta$($g'$--$r'$)\,$\approx$\,0.1\,mag. Here, we quantify the evolution of the source colour in more detail.

\begin{figure*}
  \begin{center}
    \includegraphics[angle=0,width=8.5cm]{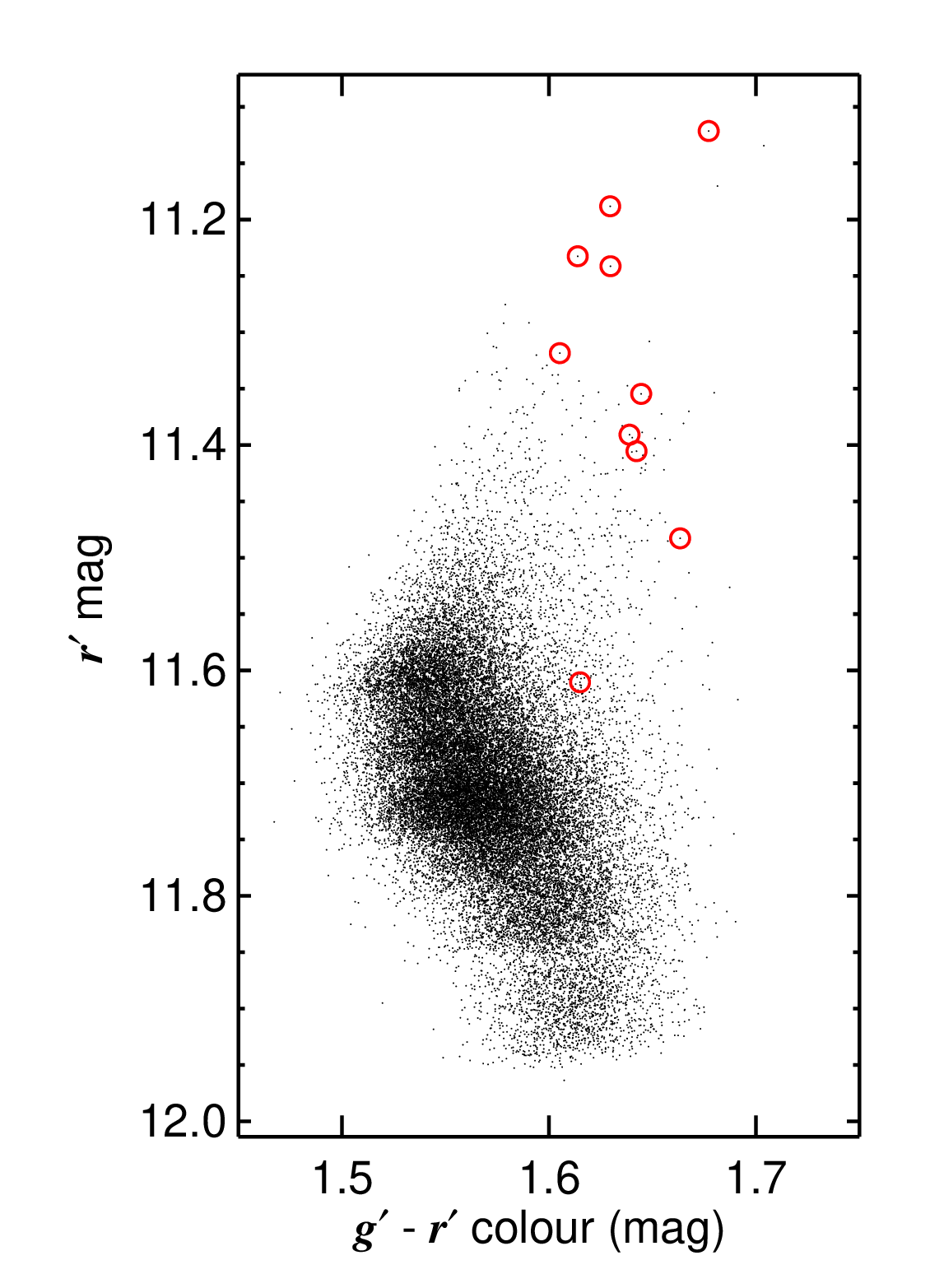}
    \includegraphics[angle=0,width=8.5cm]{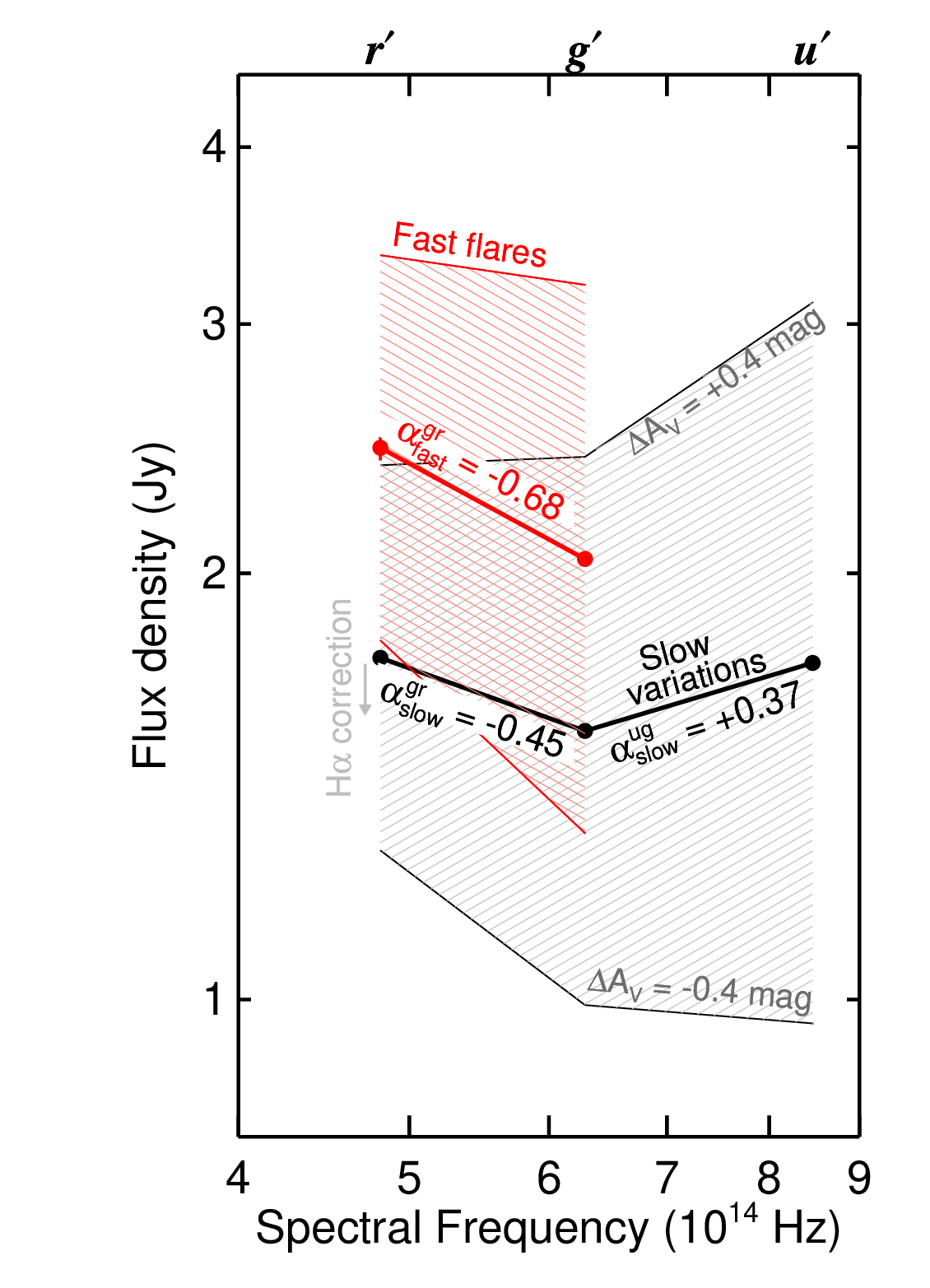}
    \caption{{\em (Left)} Colour vs. magnitude for each of the 32,780 time bins of the fast $g'$ and $r'$ second epoch light curves from June\,26. The 10 brightest, unique flares selected in Fig.\,\ref{fig:lc2} are highlighted with red circles. 
      {\em (Right)} The mean $u'g'r'$ dereddened SED of V404\,Cyg during the second epoch June\,26 observation measured at a time resolution of $\approx$\,1\,s is shown in black. This is contrasted with the mean fast $g'r'$ SED from the times of the 10 brightest, unique flares (24\,ms cycle time) in thick red. The hatched zones show the effect of 10\%\ dereddening uncertainties for both these components, and the grey arrow shows the decrease in the $r'$ flux after removal of the contribution of \ha\ to the broadband photometry. The SED of the flares alone -- i.e. the difference between the fast and the slow SEDs -- has a steep slope $\alpha^{gr}_{\rm flares}$\,$\approx$\,--1.3 (not shown for clarity).
 \label{fig:colours}}
  \end{center}
\end{figure*}

Fig.\,\ref{fig:colours} shows the $g'$--$r'$ colour vs. source brightness for every time bin during the second epoch on Jun\,26. There are two clear regimes visible. The majority of points follow a locus extending diagonally upwards to the left, showing a bluer (smaller) colour when the source increases in brightness. But above this locus, the trend is reversed, with the source becoming redder (larger $g'$--$r'$ colour) at the instants when it is brightest. These two regimes represent the slow variations and the fast flares, respectively. The colour changes by $\approx$\,0.1\,mag in both regimes. Using a Kolmogorov-Smirnov test to compare the colour distribution of the 10 brightest, unique flares selected in the previous section with the colours of all the remaining time bins yields a null hypothesis probability of 2\,$\times$\,10$^{-7}$, i.e. the two distributions differ significantly. 

The right panel of the same figure shows the mean ULTRACAM optical SED of the slow variations and of the fast flares. The former was constructed by binning all lightcurves to $\approx$\,1\,s.\footnote{More precisely, the binning factor was the closest integer multiple of the $u'$ time resolution to 1\,s which, for this night, is 1.083\,s} This effectively smooths over much of the fastest flaring, leaving slower light curves dominated by the longer variations. Observed magnitudes were dereddened assuming \av\,=\,4\,mag \citep{casares93, hynes09} and a standard Galactic reddening law \citep{cardelli89}, resulting in $u'g'r'$ SEDs for each time bin in the slow (binned) data, whose mean is shown in the figure. The fast (unbinned) $g'r'$ light curves were similarly dereddened and the total SEDs at the times of the 10 brightest flares extracted. The mean SED of these flares is also shown in the figure in red.

The SEDs of both components are found to be rising towards the red between $g'$ and $r'$, with the SED for flares rising more steeply than the slower variations. Between $g'$ and $u'$, the slope reverse and rises towards the blue for the slow variations. The SED slopes were quantified with spectral indices $\alpha$, defined as $F_{\nu}$\,$\propto$\,$\nu^{\alpha}$ and measured between the central frequencies of two filters. For the mean of the slow variations, we find $\alpha^{gr}_{\rm slow}$\,=\,--0.45\,\p\,0.09 between $g'$ and $r'$, and $\alpha^{ug}_{\rm slow}$\,=\,+0.37\,\p\,0.11 between $u'$ and $g'$. The quoted error here is the standard deviation scatter over all time bins. In contrast, at the times of the fast flares, we find a significantly steeper mean slope with $\alpha^{gr}_{\rm fast}$\,=\,--0.68\,\p\,0.09. 

It is important to note that these spectral indices can be strongly affected by systematic uncertainties on the reddening. A 10\% (0.4\,mag) error on \av\ can change the spectral slopes greatly, as shown in Fig.\,\ref{fig:colours}. However, this does not affect the fact that the fast flares are {\em redder} than the slow variations.

The total SED at the times of the fast flares will also include the contribution from the underlying slow variations, because the light curves appear to show that the two components are additive. Therefore, subtracting the total SED at the times of the fast flares from that of the slow variations would yield an estimate of the underlying variable power law (PL) associated with red flaring activity, for which we find a median $\alpha^{gr}_{\rm flares}$\,=\,--1.26 with a standard deviation of 0.38. The 10\%\ uncertainty on \av\ introduces a systematic uncertainty of 0.50 to this power law.

\subsection{Cross-correlating the optical bands}
\label{sec:ccf}

A cursory examination of Fig.\,\ref{fig:lcall} shows that the slow variations in all three optical bands are broadly similar. In addition, the sub-second flaring appears to occur fully simultaneously in both $g'$ and $r'$, as seen in the superposed light curves from June\,26 in Fig.\,\ref{fig:lc}. In order to check for any subtle inter-band lags, we computed the cross-correlation functions (CCFs) between pairs of light curves on each night. 

We began with the June\,26 CCF between the full resolution $g'$ and $r'$ light curves. The procedure for CCF measurement has been described in \citet{g10}. Since the $g'$ and $r'$ light curves have identical sampling, we simply choose a maximum time delay ($\tau_{\rm max}$) that we are interested in investigating, and compute the CCF for all time delays spanning the range of --$\tau_{\rm max}$ to +$\tau_{\rm max}$. Uncertainties can be computed by dividing the full light curve into independent sections of length 2\,$\times$\,$\tau_{\rm max}$ and computing the CCF scatter between the sections. Since Poisson noise is relatively small, we do not correct for it here. We checked the CCFs with two methods -- by interpolating the light curves to a regularly sampled grid before cross correlating, and also by use of the discrete correlation function \citep{edelsonkrolik88}.

\begin{figure*}
  \begin{center}
    \includegraphics[angle=90,width=8cm]{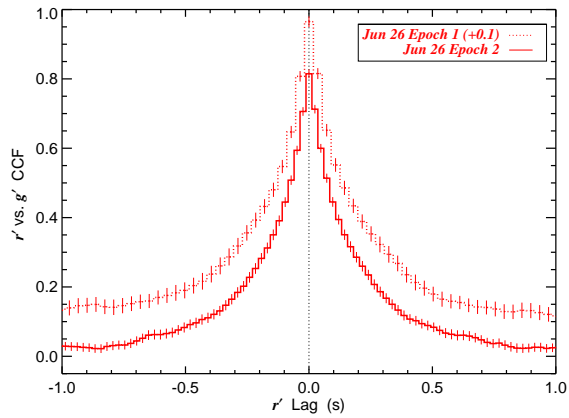}
    \caption{$g'$ vs. $r'$ CCF for the first and second epochs of the fast flaring light curves from June\,26, computed and averaged from sections of length 2\,$\times$\,$\tau_{\rm max}$\,=\,10\,s, with the uncertainties being the propagated error on the mean between sections. A small offset of 0.1 is applied to the epoch 1 CCF for clarity of display.
 \label{fig:ccf}}
  \end{center}
\end{figure*}

The result in shown in Fig.\,\ref{fig:ccf} for epochs 1 and 2 for time delays relevant for the fast flaring activity (in this case, we used $\tau_{\rm max}$\,=\,5\,s, and then zoomed in to the peak for display). The CCFs are largely symmetric and show a peak strength well over 0.8 at a peak time delay of 0\,s. In other words, there is strong correlation between the bands, and no obvious systematic time delay, as is also apparent for the fast flares from the light curves in Fig.\,\ref{fig:lc}. 

We next focused on the correlations on longer timescales. In this case, we also include the $u'$ data which has slower sampling. We rebinned the $u'g'r'$ light curves on all nights to a $\approx$\,1\,s sampling (in multiples of the $u'$ intrinsic time resolution). We found that, as we investigated longer and longer sections by increasing $\tau_{\rm max}$, an interesting asymmetry began to appear prominently, with the CCF skewed towards positive redder band time lags on all nights. However, our finite light curve durations are not long enough to investigate very long separate sections independently. As an approximation, one can compute the CCF using the entire light curve duration (duration\,=\,2\,$\times$\,$\tau_{\rm max}$), as long as one keeps in mind this caveat. In this case, we have no associated uncertainties for independent sections. 

\begin{figure*}
  \begin{center}
    \includegraphics[angle=90,width=18cm]{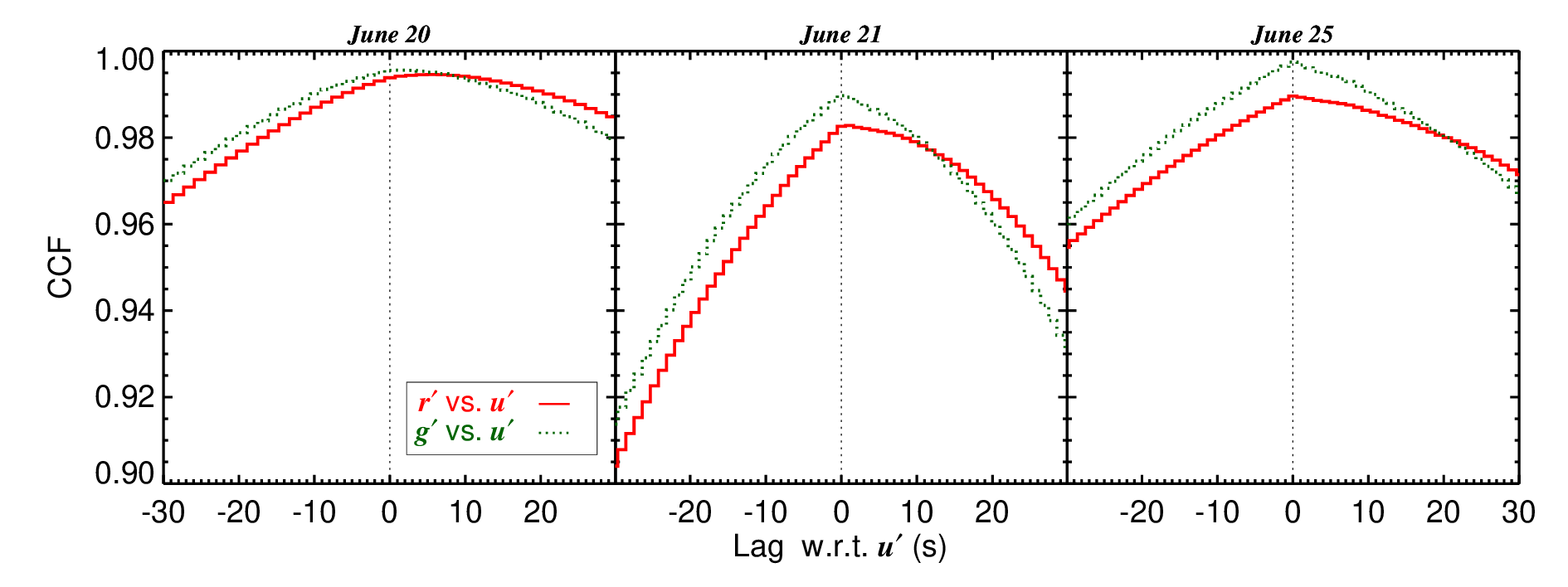}
    \caption{$g'$ vs. $u'$ and $r'$ vs. $u'$ CCFs for the nights dominated by the slow variations, computed from light curves binned to 1\,s. In this case, the entire light curve was used for computing the CCFs, and no averaging performed. 
The panels are zoomed in on the peaks, separately for each night. The lag on the x-axis is with respect to the $u'$ in all panels, with a positive value implying a lag of the redder photon with respect to $u'$. 
 \label{fig:ccfslow}}
  \end{center}
\end{figure*}

In Fig.\,\ref{fig:ccfslow}, we show these CCFs computed using the full light curve durations separately for each night. Results are shown for both the $g'$ vs. $u'$ CCFs and the $r'$ vs. $u'$ CCFs, zoomed in around the peak. Again, the CCFs are strongly correlated. The peaks are clearly much broader than the CCF relevant for the fast flares in Fig.\,\ref{fig:ccf}, as we are focusing here on the longer timescales, which are dominated by the slow variations on these nights.

But all CCFs also show the aforementioned skew towards positive lags.\footnote{This red skew due to the slow variations also contributes weakly on June\,26, and is faintly visible in Fig.\,\ref{fig:ccf}, although the CCF in this case is dominated by the fast flaring.} The skew in the $r'$ CCFs is larger than for $g'$. A red skew is indicative of the presence of red lags for the slow variations (with respect to the bluer photons). In fact, the CCFs on June\,20 peak around lags of $\approx$\,+6\,s ($r'$ vs. $u'$) and $\approx$\,+3\,s ($g'$ vs. $u'$), and on June\,21, at $\approx$\,+1\,s ($r'$ vs. $u'$). Examining various sections of the light curve individually, we indeed found evidence for changing CCF lag strengths between sections.

But the CCF lags and shapes also appear to change between the nights (although the red skew is present in all). This is likely a reflection of non-stationary behaviour of the light curves, which manifests as the CCF changes. It is clear from Fig.\,1 that the slow variations do show nightly changes, but our light curve durations do not fully sample these. This evolution needs to be investigated in detail. This is best done in Fourier space by computing the Fourier lags and coherence, and by including other long-term optical monitoring data gathered during this outburst, which we leave for future work. We note in passing that a similar skew towards positive time delays for the redder bands has also been observed in V404\,Cyg during quiescence \citep{shahbaz03}.

\subsection{Optical spectroscopy}
\label{sec:optspec}

\begin{figure}
    \begin{center}
    \includegraphics[angle=90,width=8.5cm]{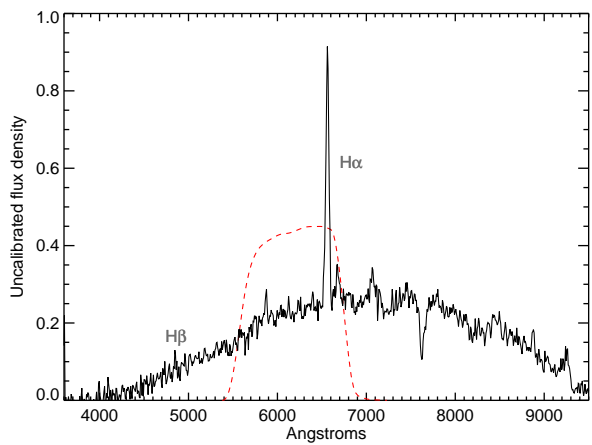}
    \caption{Optical BOOTES-2/COLORES spectrum of V404\,Cyg from 2015 June\,26. The red dashed curve shows the optical SDSS $r'$ filter response. The spectrum is not flux calibrated. 
 \label{fig:optspec}}
  \end{center}
\end{figure}

The BOOTES-2/COLORES optical spectrum is shown in Fig.\,\ref{fig:optspec}. Strong and single peaked \ha\ line in emission is the strongest feature. \hb\ appears weaker but is significant. He\,{\sc i} (5876\,${\rm \AA}$ and 6678\,${\rm \AA}$) lines are also observed, in addition to an emission line at $\approx$\,7100\,${\rm \AA}$ that might correspond to He\,{\sc i} 7065\,${\rm \AA}$.

The equivalent width (EW) of the most prominent \ha\ line was measured by fitting a Gaussian line and continuum model over the wavelength range of 6450--6640\,\AA, and we find EW\,=\,107.9\,\p\,9.3\,\AA.

In order to estimate the contribution of the line to our $r'$ photometry, the spectrum needs to be convolved with the SDSS filter response. Since our spectrum is not flux calibrated, we modelled the continuum under two extreme assumptions encompassing a broad range of possible spectral slopes: $\alpha$\,=\,+2 for a blue rising continuum, and $\alpha$\,=\,--2 for a red continuum (with $\alpha$ being the continuum slope as before). The modelled continuum is normalised to that in the observed spectrum at 6563\,\AA, so that the relative strength of the emission line is preserved at its central wavelength.

We then estimated the emission line contribution by convolving these spectra with the SDSS filter response, both with and without the addition of our fitted Gaussian emission line to the model. The result is that \ha\ contributes between 8--11\,\% to the $r'$ photometry. The magnitude and direction corresponding to a correction of 11\,\% are illustrated by the grey arrow in the right-hand panel of Fig.\,\ref{fig:colours}.

On this night, we only have a single observation in \ha. Follow-up observations on the night of June\,29 showed variations in EW(\ha) by factors of $\approx$\,2 on timescales of tens of minutes \citep{caballerogarcia15}. So the correction could potentially be even larger, though it is worth noting that the source was already significantly in decline on June\,29 and displaying different characteristics to outburst peak.

\section{Discussion}

What can we learn about the origin of the source variability during the 2015 outburst from the high temporal resolution optical lightcurves? 

Firstly, the source shows strong fluctuations in all bands and on a variety of timescales. There are at least two components to the overall variability: 1) relatively slow and smooth variations with PSD rising steeply towards low Fourier frequencies, and 2) fast sub-second flaring. Both the Fourier power and spectral colour of these components changes over the course of the outburst. The slow variations appear to be present on all nights (Fig.\,\ref{fig:lcall}). The sub-second flaring is persistent only on the last night (June\,26 UT), though it does appear sporadically on the night immediately preceding this (June\,25; Fig.\,\ref{fig:lc3}). The slower variations are bluer when brighter, whereas the sub-second flaring displays the opposite trend (Figs.\,\ref{fig:lc},\,\ref{fig:powspec_all}).

\subsection{Origin of the sub-second flaring}
\label{sec:fastdiscussion}

With an orbital period of 6.47\,days \citep{casares92}, the expected binary separation in V404\,Cyg is $\sim$\,100\,light-seconds, implying a large accretion disc size. So the fast sub-second flares seen by ULTRACAM are too speedy to be caused by standard reprocessing on any extended structures like a disc or outflowing material. Their steep red spectral slope (\S\,\ref{sec:colours}) could, instead, be consistent with optically-thin synchrotron emission (although see next paragraph for further discussion). We observe flaring on a range of timescales, with some of the sub-second flares being unresolved down to our best time resolution of 24\,ms, which would correspond to a light travel time across 500\,\gravrad\ (Gravitational radii) for a 9\,\Msun\ BH. This is consistent with the expected and inferred sizes of the jet base during hard state outbursts of some other XRBs \citep[e.g. ][]{markoff01, markoff03, casella10, g10, g11_wise, kalamkar15}. So, both the characteristic timescale and the spectral properties of the fast flares can be accounted for with optically-thin synchrotron emission from a jet. Moreover, if a single population of particles is emitting broadband radiation spanning both $g'$ and $r'$, we do not expect any time delay between the two bands, as observed (Fig.\,\ref{fig:ccf}). The fact that flaring is persistently present on June\,26 implies that the emitting jet zone is also persistent throughout the length of this observation, at least. A (quasi)stable compact jet could then be the source of these flares. 

Assuming an extinction of \av\,=\,4\,mag, we find the slope of the power law of the fast flares (above the slow variations) to be $\alpha^{gr}_{\rm flares}$\,=\,--1.3\,\p\,0.4 (\S\,\ref{sec:colours}). The distribution of jet plasma energies ($p$) is related to the observed spectral slope in the optically-thin power law as $p$\,=\,1--2\,$\alpha$ \citep{rybickilightman}, implying $p$\,=\,3.6\,\p\,0.8. This is steeper than is typical for optically-thin synchrotron in XRBs, for which $\alpha$\,$\sim$\,--0.7 (and $p$\,$\sim$\,2.5) is more common \citep[e.g. ][]{hynes03_xtej1118, g11_wise}. Steep power law slopes have been observed before (cf. the cases of XTE\,J1550--564, XTE\,J1118+480 and Swift\,J1357.2--0933 where $\alpha$\,$\approx$\,--1.3 to --1.5 were observed; \citealt{russell10, russell13_breaks, shahbaz13}). A mixture of thermal and non-thermal particle energies could potentially explain such a slope. Cyclotron-like emission has also been invoked to explain the fast optical flaring seen in other sources \citep{fabian82}. However, we again stress that reddening introduces a significant uncertainty of 0.5 in these optical power law spectral slopes, potentially making them consistent with optically-thin synchrotron at face value. 

\v404\ clearly possessed a strong radio jet during the present outburst, and our June\,26 observations were closely contemporaneous with a giant radio flare reported by \citet{trushkin15_giantflare}. This can be seen in Fig.\,\ref{fig:integral} which shows the long-term RATAN 22\,GHz radio light curve measured by \citeauthor{trushkin15_giantflare} in the top panel, and a zoom-in around our June\,26 ULTRACAM observation in the lower panel together with the 13.9\,GHz lightcurve from the AMI telescope which also caught the flare. This was the strongest radio flare of the 2015 outburst. So it is not surprising that there is a jet contribution to the optical emission at this time. It should be noted, however, that the radio spectral index shows dramatic variations in time, swinging between positive (optically-thick) and negative (optically-thin) values \citep{trushkin15_giantflare_inverted}. Some of these swings may be related to time lags between the various frequencies, as pointed out from radio and sub-mm monitoring by \citet{tatarenko15}. If so, then it is non-trivial to extrapolate between the radio and the optical bands using a simplistic power law, say. 

How much power do the optical flares carry? We corrected the $r'$ light curve of the source for interstellar extinction of \av\,=\,4\,mag using the standard reddening law of \citet{cardelli89} and computed the peak power in the strongest observed flare (lower panel of Fig.\,\ref{fig:lc2}) above the underlying slow mean as $\nu L_\nu$\,=\,4.1\,$\times$\,10$^{36}$\,erg\,s$^{-1}$, or 0.4\,\%\ of the Eddington luminosity, at the central $r'$ band wavelength of 6231\,\AA\ and for a distance of 2.4\,kpc\,\citep{millerjones09}. Assuming that the observed steep power law extends down to a frequency of 1.8\,$\times$\,10$^{14}$\,Hz where a synchrotron spectral break (\nub) was inferred to be present in the hard state of the 1989 outburst by \citet{russell13_breaks}, the integrated luminosity in the brightest flare between the assumed break and the $g'$ band is similar, at $L_{(\nu_{\rm break}-g')}$\,$\approx$\,5.4\,$\times$\,10$^{36}$\,erg\,s$^{-1}$. For comparison, we also state the integrated luminosity in the {\em mean} flare SED plotted in Fig.\,\ref{fig:colours}, which is $L_{(\nu_{\rm break}-g')}$\,$\approx$\,3.0\,$\times$\,10$^{36}$\,erg\,s$^{-1}$ above the underlying slow SED. In the jet scenario, this represents a lower limit to the integrated jet radiative power, with contributions at longer wavelengths extending to the radio at one end, and at higher frequencies extending up to the (unknown) cooling break, unaccounted for.  

Knowledge of the break frequency (\nub) can be used to constrain the magnetic field strength in the synchrotron emitting plasma. This has been done in a number of recent works assuming a single zone plasma under equipartition (e.g. \citealt{chaty11, g11_wise, russell13_breaks, tomsick15}). We do not detect an obvious break within the ULTRACAM spectral range, but the optically-thin slopes that we measure imply \nub\,$<$\,4.8\,$\times$\,10$^{14}$\,Hz (the $r'$ band central frequency). If we assume that the fast flares do originate from optically-thin synchrotron then, using the mean $r'$ flux for the bright flares (Fig.\,\ref{fig:colours}) together with Eqs.\,1 and 2 from \citet{g11_wise} for the magnetic field ($B$) and radius ($R$) of the emission zone, we find $B$\,$<$\,1.7\,$\times$\,$10^{5}$\,G and $R$\,$>$140\,\gravrad. Here, we have assumed a scale height of 1 for the optically-thin region relative to its emission radius $R$ (as defined in the appendix of \citealt{chaty11}) because the fast flares must arise from a compact region; however, we note that the above calculations are quite insensitive to the scale height and the assumption of equipartition. The above lower limit on $R$ inferred from the SED is consistent with the upper limit of $\sim$\,500\,\gravrad\ derived from the variability timescale of the unresolved flares. 

Strong fast flaring is persistent during only one of our four nights of observation. On the preceding night of June\,25, it is sporadic and short-lived (Fig.\,\ref{fig:lc3}). So it is clearly much more transient as compared to the slower variations which persist on all nights. 
Fast flaring unresolved to a time resolution of $\sim$\,1--2\,s was also found to be brief and sporadic by \citet{hynes15_atel2} and \citet{terndrup15}, for periods of up to $\sim$\,25\,mins. 
If the source of this sporadic variability were also the compact jet, this would imply strong changes in the compact jet spectrum on timescales as short as $\sim$\,100\,s, as a result of which the rapidly variable optical synchrotron component appears and disappears on these times. Variations in the location of the break frequency have been previously reported in the infrared SED of GX\,339--4 \citep{g11_wise} sampled at timescales of $\sim$\,90\,mins and less. In addition, it is noteworthy that V404\,Cyg itself has also been found to display dramatic swings in its radio jet spectral index during quiescence on timescales of tens of minutes \citep{rana15}. For completeness, we note that we cannot rule out a transient clearing (during these short periods on June\,25, and for the full duration of our observation on June\,26) of an otherwise persistent screen of absorbing matter, which then reveals fast variability in the inner source regions. 

We note that our observations were also closely contemporaneous with some of the strongest X-ray flaring activity seen in the 2015 outburst of \v404. This is illustrated in Fig.\ref{fig:integral} which shows the publicly available long-term 25--200\,keV X-ray lightcurve from the IBIS/ISGRI instrument on board the \integral\ satellite \citep{kuulkers15}. The zoom-in on June\,26 shows that the ULTRACAM observation coincided with the penultimate strong hard X-ray flare (at $\sim$\,MJD\,57199.2). The next X-ray flare (centred around $\sim$\,MJD\,57199.7) was the very brightest flare of the entire 2015 outburst, following which the source began its decline to quiescence \citep{segreto15, ferrigno15, martin-carrillo15, sivakoff15}. In contrast, the source X-ray count rate during at least two of our prior observations (June\,21 and 25) was fainter by at least an order of magnitude than on June\,26. There are no \integral\ observations from the first night of June\,20. So it may be that there was strong, fast flaring on June\,26 in the optical because that was when the source was particularly active in radio and in X-rays. One may speculate that we are observing jet activity related to the ejections in a final, massive flare and source shut-off, but this needs further investigation.

\subsection{Origin of the slow variations}
\label{sec:slowdiscussion}

The origin of the slow variations is less clear. They differ from the sub-second flaring in several aspects. Their characteristic timescale of several hundred seconds is, of course, orders of magnitude longer than the fast flares. In addition, they also differ in terms of their colours (being bluer when brighter, with stronger variability power in the bluer bands; Fig.\,\ref{fig:powspec_all}) and also appear to be much more persistent over the entire set of ULTRACAM observations (as well as other observations; \citealt{hardy15_atel, hynes15_atel1, hynes15_atel2, terndrup15, wiersema15, scarpaci15_atel}). Moreover, there is evidence for positive time delays and skewed CCFs in the sense of the redder bands lagging the blue (Fig.\,\ref{fig:ccfslow}). The long timescales and red lags argue against optically-thin emission from a single-zone compact jet, whose emission is likely to vary on much faster timescales and be correlated between the bands. 

\begin{figure*}
  \begin{center}
    \includegraphics[angle=90,width=12cm]{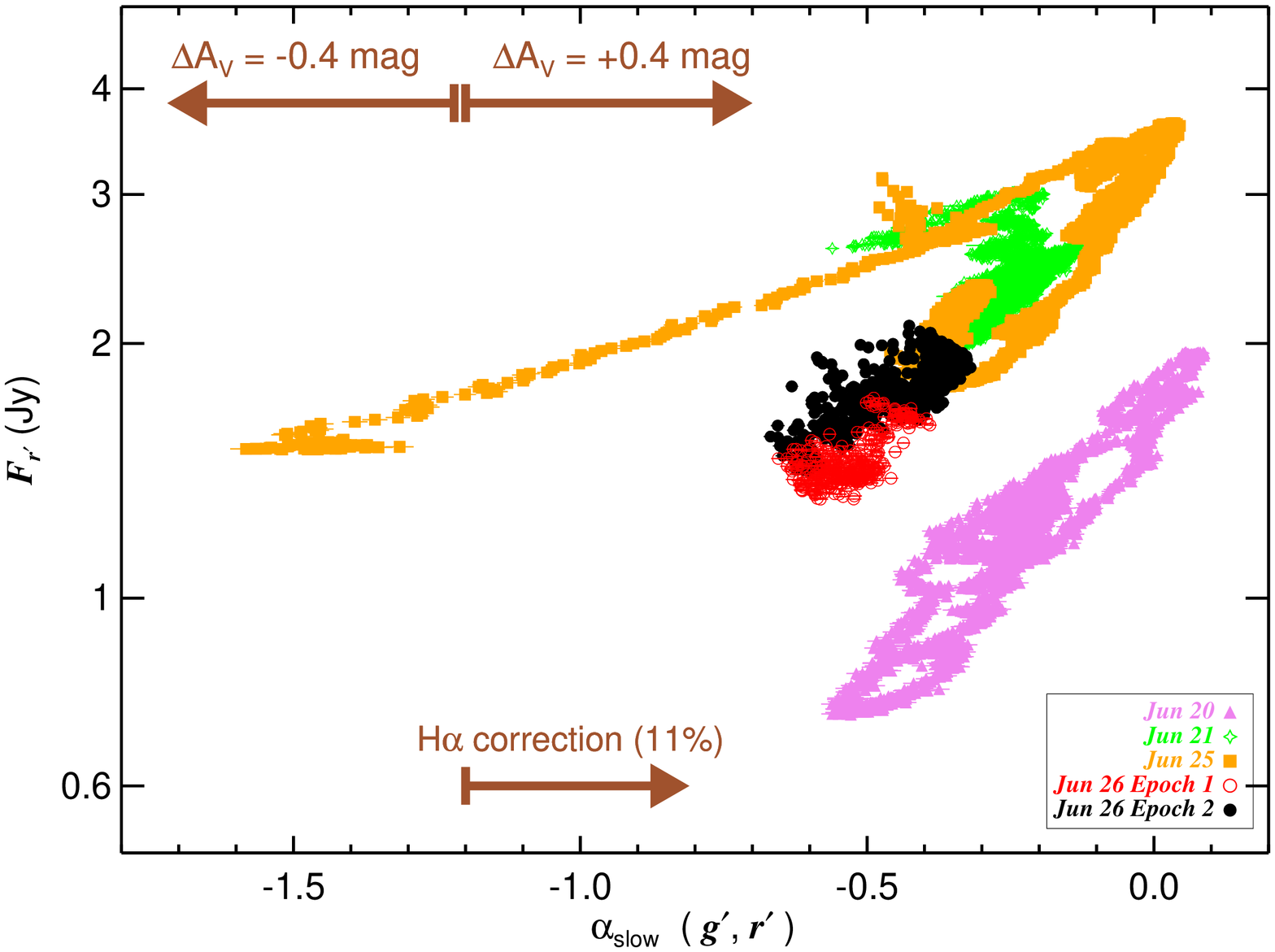}
    \caption{Flux vs. spectral slopes ($\alpha_{\rm slow}$ between $g'$ and $r'$) for all ULTRACAM observations for the slow variations, computed from light curves binned to $\approx$\,1\,s time resolution. The direction and magnitude of systematic corrections due to reddening uncertainties, and correction for the contribution of \ha\ (as estimated on June\,26), are denoted by the arrows. 
 \label{fig:colour_mag_allnights}}
  \end{center}
\end{figure*}

The SED spectral slopes are certainly bluer than those of the fast flares on the night of June\,26 (Fig.\,\ref{fig:colours}). What about the spectral slopes on the other nights? Fig.\,\ref{fig:colour_mag_allnights} shows 
$\alpha_{\rm slow} (g', r')$ as a function of flux for the slow variations on all nights. This is similar to the colour--flux plot for June\,26 (Fig.\,\ref{fig:colours}), with a larger (positive) spectral slope $\alpha$ here being equivalent to a smaller (bluer) $g'$--$r'$ colour. 
There are clear and strong variations in spectral slopes on all nights, with a full range of variations of $\Delta \alpha_{\rm slow}$ of $\approx$\,1.7. 
Yet, there are underlying patterns to these variations. In particular, there appears to be a diagonal locus of variations in the $\alpha_{\rm slow}$--flux plane, with slope increasing as the source becomes brighter. The slope of this variation is approximately similar on all nights, although the first night of June\,20 shows a systematic offset from the other nights.\footnote{This offset corresponds to a flux offset of $\approx$\,10\%\ (either a decrease in $g'$ or an increase in $r'$ by this factor would shift the locus for this night to be in agreement with the others, though we could not identify an obvious source that could introduce such as shift).} Furthermore, the night of June\,25 shows an additional evolution off towards the lower left in this plane towards steep negative slopes as the source dims (this occurs during the sharp flux drop at the end of the observation on this night; Fig.\,\ref{fig:lcall}). 

Any model for these slow variations must be able to account for these dramatic spectral slope variations. However, we also caution that the absolute values of $\alpha_{\rm slow}$ are subject to significant uncertainties. The dereddening uncertainties of 10\%\ discussed in section\,\ref{sec:colours} introduce a systematic shift of $\Delta \alpha_{\rm slow}$\,=\,0.5, denoted by arrows in the figure (and also shown as the shaded regions in the SED of June\,26 in Fig.\,\ref{fig:colours}). Furthermore, 
the \ha\ emission line contribution was found to be 11\%\ in analysis of our optical spectrum quasi-simultaneous on June\,26. Removing this can yield the {\em continuum} spectral slope free of \ha, which is larger (bluer) than $\alpha_{\rm slow} (g', r')$ by $\Delta \alpha_{\rm slow}$\,$\approx$\,+0.39. The magnitude of this correction is also shown in Figs.\,\ref{fig:colours} and \ref{fig:colour_mag_allnights}. Whereas other emission lines such as \hb\ would affect the other bands, their effect is expected to be smaller. However, \ha\ is known to be strongly variable during the outburst \citep[e.g.][]{wagner15_atel, munozdarias15_atel, scarpaci15_atel, caballerogarcia15}, and a single, average correction is undoubtedly an over-simplification. Given all these potential systematics, we are cautious about drawing any detailed conclusions about the slow variations, and restrict ourselves to a more qualitative discussion below.

\begin{figure*}
  \begin{center}
    \includegraphics[angle=90,width=18cm]{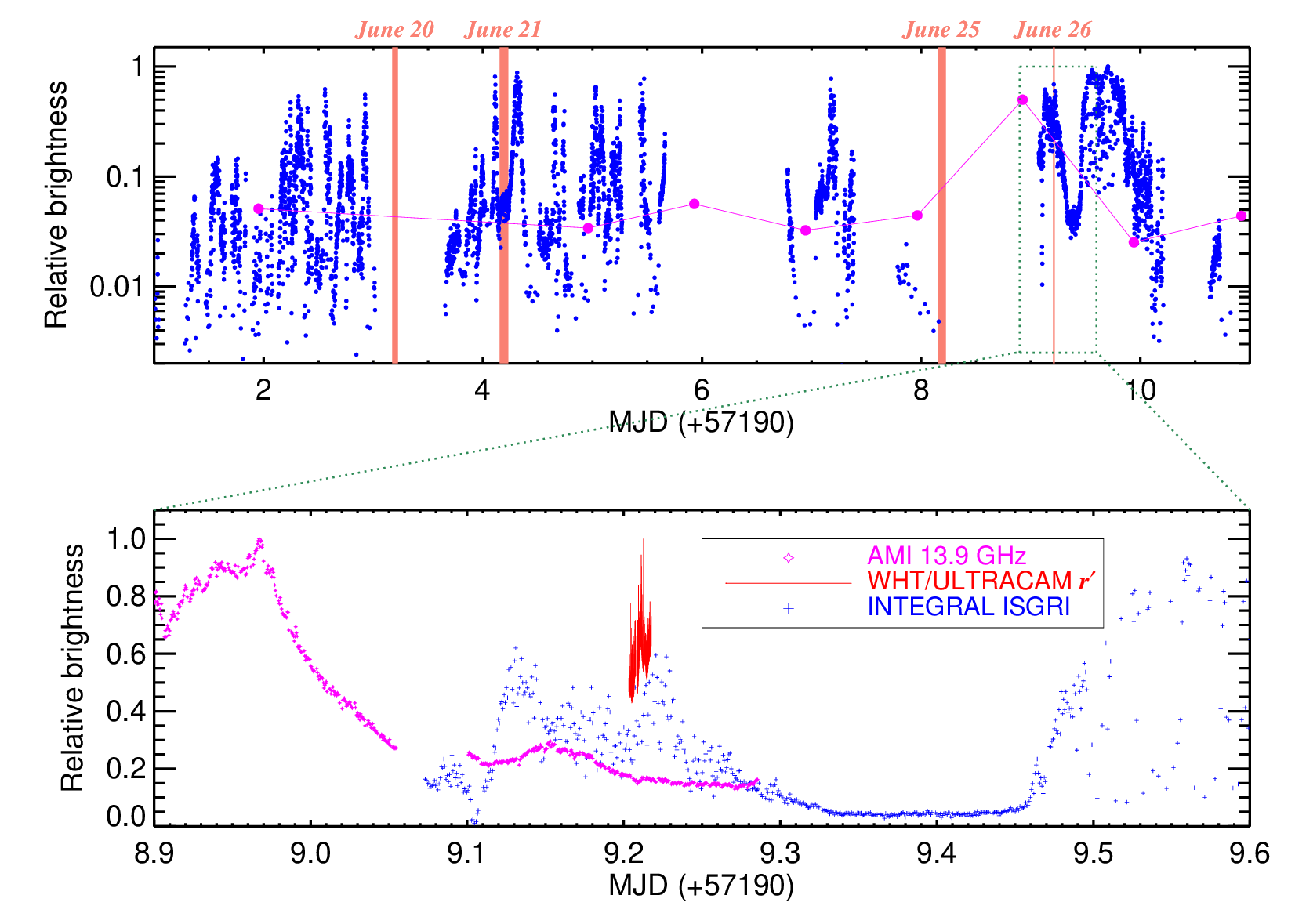}
\caption{The top panel shows the \integral\ 25--200\,keV light curve (based upon Off-line Scientific Analysis\,v.10.1) for the first $\approx$\,10\,days of the June\,2015 outburst of V404\,Cyg from the IBIS/ISGRI instrument (blue; \citealt{kuulkers15}), overlaid with the 22\,GHz radio light curve from the RATAN-600 telescope (magenta; \citealt{trushkin15_giantflare}). The times of the ULTRACAM observations are highlighted by the orange shaded strips. The bottom panel focuses on the last night around the ULTRACAM data (shown in red). In addition, the 13.9\,GHz light curve from AMI overlapping with our ULTRACAM coverage is shown in magenta. The ULTRACAM sub-second flaring was caught when the source was very active in both radio and X-rays, in between some of the strongest multiwavelength flares over the entire outburst. 
 \label{fig:integral}}
  \end{center}
\end{figure*}

Can reprocessing explain the ULTRACAM light curves? Using X-ray observations from the \swift\ satellite carried out simultaneously with the June\,21 ULTRACAM run, \cite{g15_atel2} found a close correlation between the slow optical and X-ray variations, with an optical delay on the order of tens of seconds. Specifically, the correlation was found during the time when the oscillatory pattern with a characteristic timescale of a few hundred seconds was present on this night. Such a delay would be consistent with a reprocessing origin for the expected size of the accretion disc in V404\,Cyg. If so, the oscillatory pattern could potentially result from X-ray reprocessing.\footnote{Which is not to say that other components in the optical light curves are unrelated to reprocessing, since the ULTRACAM/\swift\ coverage was only simultaneous during this period.} 
But this oscillatory pattern is not obviously present on the other nights, suggesting that any contribution from reprocessing may be changing between the nights. This is supported by the changing nightly optical inter-band time lags found in our optical data (Fig.\,\ref{fig:ccfslow}). \citet{rodriguez15} also found evidence of a varying reprocessing contribution from the \integral\ data, with the optical lag varying between $\sim$0 to 20--30\,mins. These long lags are much larger than the longest possible reprocessing (light-travel) time for the size of the binary. We also note that the optical variations across the four nights of our observations are relatively mild as compared to the X-ray variability which is orders of magnitude larger (see Fig.\,\ref{fig:integral}). In other words, the optical variations do not follow the $L_{\rm X-ray}^{1/2}$ trend expected from standard disc reprocessing \citep{vanparadijsmcclintock94}. 

It is also unclear whether scattering and absorption due to intervening matter plays an important role in explaining the optical variations. X-ray observations suggest that absorption due to a wind may be high enough to hide the central regions even at high energies \citep{king15, kuulkers15a, motta15_atel}, in which case any optical photons from the central regions would also not escape. But if electron scattering were to dominate, variability is expected to be wavelength-independent, which is not what we observe in the optical. Changing dust reddening in a wind may be a possibility, but in this scenario, the apparently stable loci of the spectral slope evolution (Fig.\,\ref{fig:colour_mag_allnights}) would require reddening levels to be fine-tuned to the flux changes across the different nights, which seems unlikely. Moreover, the naive expectation would be to see much stronger optical variations (perhaps orders of magnitude in flux) in response to a clumpy wind which is dense enough to significantly absorb X-rays. Despite the impressive variability observed, the ULTRACAM data do not show such strong variations, although we cannot rule out their presence on other nights when we did not observe. 

Using simultaneous optical/radio monitoring, \citet{mooley15} found that the strong and slow optical variations correlate with variations seen in the radio by the AMI telescope, suggesting a non-thermal contribution to the emission process for the slow optical variations. Moreover, flux-dependent spectral slope variations similar to those that we see in Fig.\,\ref{fig:colour_mag_allnights} have also been linked to synchrotron jet emission in other sources \citep[e.g. observations of XTE\,J1550--564 by ][]{russell10}. The magnetic field associated with the slow variations is likely to be much lower than the compact fast flares (\citealt{tatarenko15} estimate a field strength of $\sim$\,few Gauss). The fact that $\alpha_{\rm slow}$ for the slow variations is slightly flatter than the $\alpha_{\rm fast}$ values measured for the fast flares on June\,26 (see Fig.\,\ref{fig:colours}) could be a result either of mixing of multiple emission components (non-thermal radiation and reprocessing, say), or a differing distribution of particle energies between the slow variations and the fast flares.

The ultimate origin of these variations appears to be complex, and we do not attempt to model them further here. But we emphasise that any model for these slow variations must account for their rather long characteristic timescales. The light travel time for variations spanning $\sim$\,100--1000\,s is similar to, or even larger than, the maximal extent of the disc of 10$^{12}$\,cm (or $\sim$\,7\,$\times$\,10$^5$\,\gravrad) inferred by \citet{zycki99} during the 1989 outburst. So if the variations result from cyclotron or optically-thin synchrotron in an extended, perhaps multi-zone, plasma (e.g. ejected blobs), these ejecta ought to be very large, with sizes commensurate with the binary itself. Instead, the size of an emission zone in a standard viscous thin disc varying on $\sim$\,1000\,s timescales would be $\sim$\,3\,$\times$\,10$^8$\,cm, or 250\,\gravrad\,($M$/9\,\Msun)\,$(\alpha/0.1)^{-1}$\,$(h/R/0.1)^{-2}$, where $\alpha$ here represents the viscosity parameter and $h/R$ is the relative disc height scale (though the observed spectral slopes do not support emission from a viscous disc alone, as already discussed). 
If the slow variations are instead associated with some form of thermal instability timescale, a larger size scale of $\sim$\,5000\,\gravrad\,($M$/9\,\Msun)\,$(\alpha/0.1)^{-1}$ is possible \citep{accretionpowerastrophysics}. Accretion rate variations on these scales, which are much smaller than the overall binary size, could then manifest in the optical. For instance, synchrotron emission in a steady jet in which fast variations are \lq smoothed\rq\ out over multiple emission zones, with only the more substantial longer variations remaining, is one plausible scenario to explain our observations.

Finally, we note that no strong quasi-periodic oscillations (QPOs) are found in our PSDs, which are instead dominated by broadband noise features. This appears to be in contrast to QPOs seen in the optical and infrared PSDs of other XRBs during hard state outbursts \citep[e.g. ][ and references therein]{g09_rmsflux, kalamkar15}, although we note that we cannot unambiguously rule out the presence of low frequency QPOs given the short duration of our lightcurves. In particular, the night of June\,21 shows the oscillatory patterns of variability on timescales of $\sim$\,200--500\,s (Fig.\,\ref{fig:lcall}), and found to be correlated with simultaneous X-ray variations by \citet{g15_atel2}. There is excess power in the PSDs on these timescales (Fig.\,\ref{fig:powspec_all}), but no sharp QPO-like feature stands out immediately. So the presence of any precessing hot flow, as inferred in other sources \citep[e.g. ][]{ingramdone11}, does not stand out in our observations. One caveat in this discussion is that our optical light curves may be diluted by emission lines (in particular \ha\ and \hb\ which lie in the $r'$ and $g'$ bands, respectively), and extended line emission zones may smooth over underlying continuum QPO variability.

\subsection{Multi-component optical variability in V404\,Cyg and other XRBs} 

Multicomponent optical variability on short timescales, including potential contributions from a hot flow, accretion disc reprocessing, and (in several cases) a jet, appear to be a common feature in many XRBs during the hard state. 
Strong and fast sub-second flaring behaviour has been seen in a number of XRBs during the hard state, in particular GX\,339--4 \citep{motch82, imamura90, steiman-cameron97, g08, fender97, g10, casella10} and XTE\,J1118+480 \citep{kanbach01,malzac04}, where it has been attributed to synchrotron emission from a jet with increasing dominance towards the red. A steep spectral slope in the optical has also been observed in XTE\,J1118+480 during the hard state with strong jet activity (e.g. \citealt{hynes03_xtej1118}). 

On intermediate ($\sim$seconds to mins) timescales, V4641\,Sgr and Swift\,J1753.5--0127 have shown pronounced optical variations \citep{uemura02_v4641sgr, durant09}. Whereas Swift\,J1753.5--0127 appears to be dominated by emission from a hot flow in the optical \citep{durant09, veledina11}, those in V4641\,Sgr remain unclear, though a non-thermal origin appears to be very likely \citep{uemura04}. Finally, infrared variations on timescales of minutes to hours have been extensively studied in GRS\,1915+105 and associated with synchrotron emission \citep{fender97, eikenberry98}. Their characteristic timescales appear to be quite similar to the slow optical variations in V404\,Cyg. Infrared/X-ray cross-correlations have confirmed this, and find that the infrared emission zone can be associated with a synchrotron-emitting plasma on a large scale of $\sim$few\,$\times$\,10$^{11}$\,cm \citep{lassocabrera13}. 

The fast optical flares seen in GX\,339--4 were somewhat stronger than observed in V404\,Cyg, with several of the $\sim$\,100\,ms $r'$ and $g'$ flares during the 2007 hard state observation of GX\,339-4 exceeding a factor of 2 above mean \citep{g10}. The peak optical flare luminosities in that outburst reached $\sim$\,10$^{36}$\,erg\,s$^{-1}$, of the same order as observed for V404\,Cyg here. In section\,\ref{sec:colours}, we have shown that subtracting the SED at the times of the fast flares from that of the slow variations on June\,26 gives a median $\alpha^{gr}_{\rm flares}$\,=\,--1.3\,\p\,0.4 as an estimate of the variable PL whose contribution is reddening the flare SEDs. 
This slope is steeper than seen in GX\,339--4 \citep{g11_wise}, but is similar to those seen in XTE\,J1118+480 and XTE\,J1550--564 during outburst \citep{russell13_breaks}. 

On the other hand, the optical/X-ray CCF strength appears to be much stronger in V404\,Cyg \citep{g15_atel2}. As already discussed, this cannot be entirely ascribed to reprocessing, although there is strong \ha\ emission which is likely to dilute the true $r'$ band continuum variability \citep[e.g.][]{wagner15_atel, munozdarias15_atel, scarpaci15_atel}. With an orbital period approximately four times longer than that of GX\,339--4, V404\,Cyg's accretion disc is also expected to be larger, which sould (in principle) result in stronger reprocessing and also dilute the strength of any synchrotron jet flares.

\section{Conclusions}

We have analysed high temporal resolution optical multi-band ($u'g'r'$) light curves obtained with ULTRACAM during the June\,2015 outburst. We are able to probe variations over approximately five orders of magnitude in Fourier frequency. Our main results are summarised below.

\begin{enumerate} 

\item The data show a diverse range of variations on all timescales, but reveal at least two prominent variability components: (1) \lq slow variations\rq\ with characteristic timescales of hundreds of seconds, and (2) \lq fast sub-second flares\rq\ (section\,\ref{sec:lc}, Figs.\,\ref{fig:lcall},\,\ref{fig:lc},\,\ref{fig:lc2}). 

\item These components differ in their colours and time lags. The fast flares show red colours at peak, whereas the slow variations are stronger in the bluer bands. (section\,\ref{sec:colours}, Figs.\,\ref{fig:lc}, \ref{fig:colours}). 

\item The multi-band PSDs are dominated by the slow variations and rise steeply towards low Fourier frequencies on all nights, but show a dramatic increase in the $\sim$\,1\,Hz variability power by a factor of $\approx$\,100 on the last night of June\,26. This night clearly shows the reversal of the dominant source of variability power above and below $\sim$\,0.01\,Hz between the faster red variations and slower blue variations (section\,\ref{sec:psd}, Fig.\,\ref{fig:powspec_all}). 

\item The fast flares display complex flare profiles (Figs.\,\ref{fig:lc2}, \ref{fig:unresolvedflares}), but show no time lags between $g'$ and $r'$ down to the best time resolution of 24\,ms (section\,\ref{sec:fastflares}, Fig.\,\ref{fig:ccf}). In contrast, the cross-correlation functions of the slow variations are indicative of underlying, weak red lags of a few seconds (section\,\ref{sec:ccf}, Fig.\,\ref{fig:ccfslow}). 

\item We interpret these observables in terms of an optically-thin synchrotron origin for the fast flares having short characteristic timescales and high optical luminosities $\sim$\,10$^{36}$\,erg\,s$^{-1}$ (section\,\ref{sec:fastdiscussion}). The spectral slopes of the fast flares after removing the underlying slow variations is $\alpha^{gr}_{\rm flares}$\,$\sim$\,--1.3, which is steeper than typical for optically-thin synchrotron from a compact jet alone, and could be indicative of a mixture by a thermal particle distribution. However, including systematic dereddening uncertainties, this slope could also be consistent with $\alpha^{gr}_{\rm flares}$\,$\sim$\,--0.8. 

\item Under the compact jet scenario, we can place limits on the magnetic field strength at the synchrotron emission zone of $B$\,$\ltsim$\,2\,$\times$\,10$^5$\,G, and a zone size $R$\,$\gtsim$\,140\,\gravrad. If the fastest flares arise within this zone, the variability timescale of the unresolved flares of $<$\,24\,ms implies $R$\,$\ltsim$\,500\,\gravrad\ (section\,\ref{sec:fastdiscussion}). 

\item These fast flares are prominent and persistent on the night when the source reached the peak of its outburst in terms of its observed X-ray flux, and also showed the brightest radio flares (Fig.\,\ref{fig:integral}). They also appear sporadically on the preceding night with short durations of $\sim$\,30\,s (Fig.\,\ref{fig:lc3}), which places constraints on the timescales over which the compact jet spectrum can change drastically, appearing and disappearing in the optical on timescales of order $\sim$\,minutes or less (section\,\ref{sec:fastdiscussion}) -- assuming that they do, indeed, arise from a compact jet, and their appearance/disappearance is not controlled by other effects such as changing line-of-sight absorption/reddening, say. 

\item On the other hand, the origin of the slow variations is far from clear. X-ray reprocessing likely plays a role in some of these variations but appears unlikely to be their sole driver, and there is suggestive evidence of a non-thermal contribution to the slow variations also (section\,\ref{sec:slowdiscussion}; cf. reports by other authors of changing reprocessing contributions with time, and of correlations between radio and optical flaring). 

\item We find that the optical spectral slopes can change rapidly but evolve according to quasi-stable loci across all nights (Fig.\,\ref{fig:colour_mag_allnights}). However, we caution that dereddening uncertainties and the contribution of \ha\ can strongly affect the optical photometric spectral slopes (section\,\ref{sec:colours}, \ref{sec:optspec}, \ref{sec:slowdiscussion}). 

\item No obvious QPOs are found in the optical light curves, though we note the presence of an oscillatory pattern on intermediate timescales of minutes on one night, reported (in a previous work) to be correlated with X-ray variations (section\,\ref{sec:slowdiscussion}; Fig.\,\ref{fig:lcall}). 

\end{enumerate}

\vspace*{1cm}
\noindent
There is a wealth of high-quality optical data (especially from small telescopes) during the 2015 outburst, examination of which should shed more light on the nature of the slow variations. In addition, this outburst galvanised much of the XRB community to coordinate multiwavelength observational efforts \cite[e.g. ][]{knigge15}, resulting in strictly simultaneous periods of coverage over much of the electromagnetic spectrum. Cross-correlation of fast optical and X-ray light curves \citep[e.g. ][]{kanbach01, g08, durant08}, and investigation of infrared and radio light curves (Dallilar et al. in prep.) may help to answer many of the questions raised herein.

\section{Acknowledgements}

P.G. acknowledges funding from STFC (ST/J003697/1), and thanks C. Done, P. Uttley, S.D. Connolly and R.E. Firth for discussions. Part of this research was supported by the UK-India UKIERI/UGC Thematic Partnership grant UGC 2014-15/02. \ultracam\ is supported by STFC grant PP/D002370/1. D.A. thanks the Royal Society for support. T.S., J.C. and T.M.-D. were supported by the Spanish Ministry of Economy and Competitiveness (MINECO) under the grant AYA2013-42627. T.R.M. acknowledges STFC (ST/L000733/1). J.V.H.S. acknowledges support via studentships from CONACyT (Mexico) and the University of Southampton. J.C. also acknowledges support by DGI of the Spanish Ministerio de Educaci\'on, Cultura y Deporte under grant PR2015-00397. P.C. acknowledges support by a Marie Curie FP7-Reintegration-Grants under contract no. 2012-322259. We acknowledge with thanks the variable star observations from the AAVSO International Database contributed by observers worldwide and used in this research. The report by the anonymous reviewer is acknowledged. 

\section{Addendum}

Post-submission of our paper, several new preprints and publications on the outburst have become public. Their key results relevant to our work are briefly mentioned here. With regard to the origin of the optical variability, \citet{kimura16} present an analysis of the public optical photometric data gathered by small telescopes covering the majority of the bright part of the outburst, and show that the slow variations have distinct parallels with the (X-ray) variability modes identified in GRS\,1915+105. They conclude that limit cycle oscillations in the inner disc generate X-ray variations, which are then reprocessed on the outer disc and generate the slow optical variability observed (\citealt{bernardini16} similarly interpret the pre-outburst source evolution in terms of a viscous thermal instability). The $\sim$\,minute--timescale oscillatory pattern that we observe on June\,21 is similar to what \citeauthor{kimura16} term the \lq heartbeat\rq\ oscillation in their light curves. In contrast to a reprocessing scenario, \citet{marti16} tentatively interpret the source behaviour in terms of synchrotron radiation from expanding plasmons, again drawing a parallel to GRS\,1915+105. These authors also report the presence of redder band lags with respect to bluer ones, with the lags being around $\sim$\,1\,min (longer than what we report), but changeable and difficult to pin down. The properties of jet models have been further investigated by \citet{jenke16} and \citet{tanaka16}. Based upon modelling of \fermi\ Gamma Ray Burst Monitor observations, \citeauthor{jenke16} propose that X-rays from a jet could be seeding Compton-upscattering to the Gamma ray band, similar to the conclusions drawn earlier by \citet{roques15} based upon \integral\ analysis. \citet{tanaka16} find a low linear optical polarisation degree intrinsic to the source, and suggest either a disc or optically-thick synchrotron origin for the optical emission. They also derive constraints on the physical parameters of the jet ($B$ field $\sim$\,10$^5$\,G and emission region size $\approx$\,10$^8$\,cm) which are of the same order of magnitude as the limits that we present in section\,\ref{sec:fastdiscussion}. Finally, \citet{radhika16} study the \swift/XRT source spectra and conclude that absorption features and a variable Fe line point to a wind origin. 

All the above works, and their (sometimes) disparate conclusions, emphasise the complexity of the present outburst of V404\,Cyg. It is important to keep in mind that most of the controversy centres on the variable component that we refer to as the \lq slow variations\rq. 
In contrast, the study of the sub-second variations is unique to our work, and (as we have discussed) their properties all point to a non-thermal origin, and are consistent with arising in a compact jet. 

It is worth noting the recent launch of the \astrosat\ mission by the Indian Space Research Organisation \citep{astrosat}. Its Large Area X-ray Proportional Counter instrument can provide exquisite sensitivity at high time resolution. Simultaneous fast timing capability in the ultraviolet is provided by the Ultra Violet Imaging Telescope \citep{uvit}. Given the fast optical variability that we have presented in our work, this combination of instruments can be very effective for the study of future XRB outbursts. The approximate 30\,year timescale between outbursts of V404\,Cyg suggests that it may be too optimistic to hope for another outburst from it any time soon, although we note that in December\,2015, the source did undergo a mini outburst (possibly a secondary to the June outburst) which was followed up by several telescopes \citep[e.g. ][]{barthelmy15_v404wakeup,lipunov15_v404wakeup,jenke15_v404wakeup,trushkin15_v404wakeup,hardy16_v404wakeup}, and whose detailed study will be important. And, as already noted, V404\,Cyg is well known for its variability even in quiescence \citep[e.g. ][]{shahbaz03, bernardinicackett14, rana15}. So future observations of V404\,Cyg with \astrosat\ could still prove useful.

\setcounter{figure}{0}
\makeatletter 
\renewcommand{\thefigure}{A\@arabic\c@figure}
\makeatother
\setcounter{section}{0}
\makeatletter 
\renewcommand{\thesection}{A\@arabic\c@section}
\makeatother

\section{Appendix}

\subsection{Flux calibration}

Ideally, one would use the comparison star for photometry. However, V404\,Cyg was the brightest object in the field of view during outburst, especially in the red, and the comparison stars used are not photometric standards. Therefore, the ULTRACAM flux calibration is based upon zeropoints derived from measurements of two photometric standard stars on two nights during the week of observation. An independent, approximate test of the flux calibration is possible using AAVSO\footnote{Kafka, S., 2015, Observations from the AAVSO International Database, {\tt http://www.aavso.org}} data, which had good overlap with the ULTRACAM observations on the two nights of June\,20 and 25. We also checked against comparison star photometry and computed resulting systematic uncertainties on the flux calibration, all of which is described below. 

The photometric standards were HD\,121968 observed on June\,24 in $u'$, $g'$ and $r'$, and GCRV\,8758 observed in $u'$, $g'$ and $z'$ on June\,26 (we do not use the $z'$ calibration herein). The zeropoints measured are 25.02 ($u'$), 26.96 ($g'$) and 26.59 ($r'$), respectively, on June\,24, and 25.09 ($u'$) and 26.95 ($g'$) on June\,26. These zeropoints apply to an incident rate of 1 electron per second, with gain factors of 1.16, 1.11 and 1.19 in $u'g'r'$, respectively. We assumed zeropoints of 25.09 ($u'$), 26.95 ($g'$) and 26.59 ($r'$) throughout our work. All observations were carried out not far off zenith, at airmass $<$\,1.1. Airmass corrections were applied for each night, using $u'g'r'$ extinction coefficients of 0.49, 0.16 and 0.07, respectively. 

We checked our calibration by computing the magnitudes of the comparison stars. The comparison star was URAT1\,620-473723 \citep{zacharias15} at RA\,=\,20:24:07.181, Dec\,=\,+33:50:51.66 on June\,20, which is also identified as star \lq C1\rq\ of \citet{udalski91}. Hereafter, this is referred to as \lq comparison star 1\rq. On the remaining nights, the comparison star was URAT1\,620-473466 at RA\,=\,20:23:56.44, Dec\,=\,+33:48:16.9 (\lq comparison star 2\rq). The weather conditions were good on all nights. Comparing the nightly median count rates of comparison star 2 during June\,21--26, we find agreement to within 4\%, 0.5\% and 1\% (standard deviation) in $u'$, $g'$ and $r'$, respectively.

\begin{figure*}
    \begin{center}
    \includegraphics[angle=0,width=8.5cm]{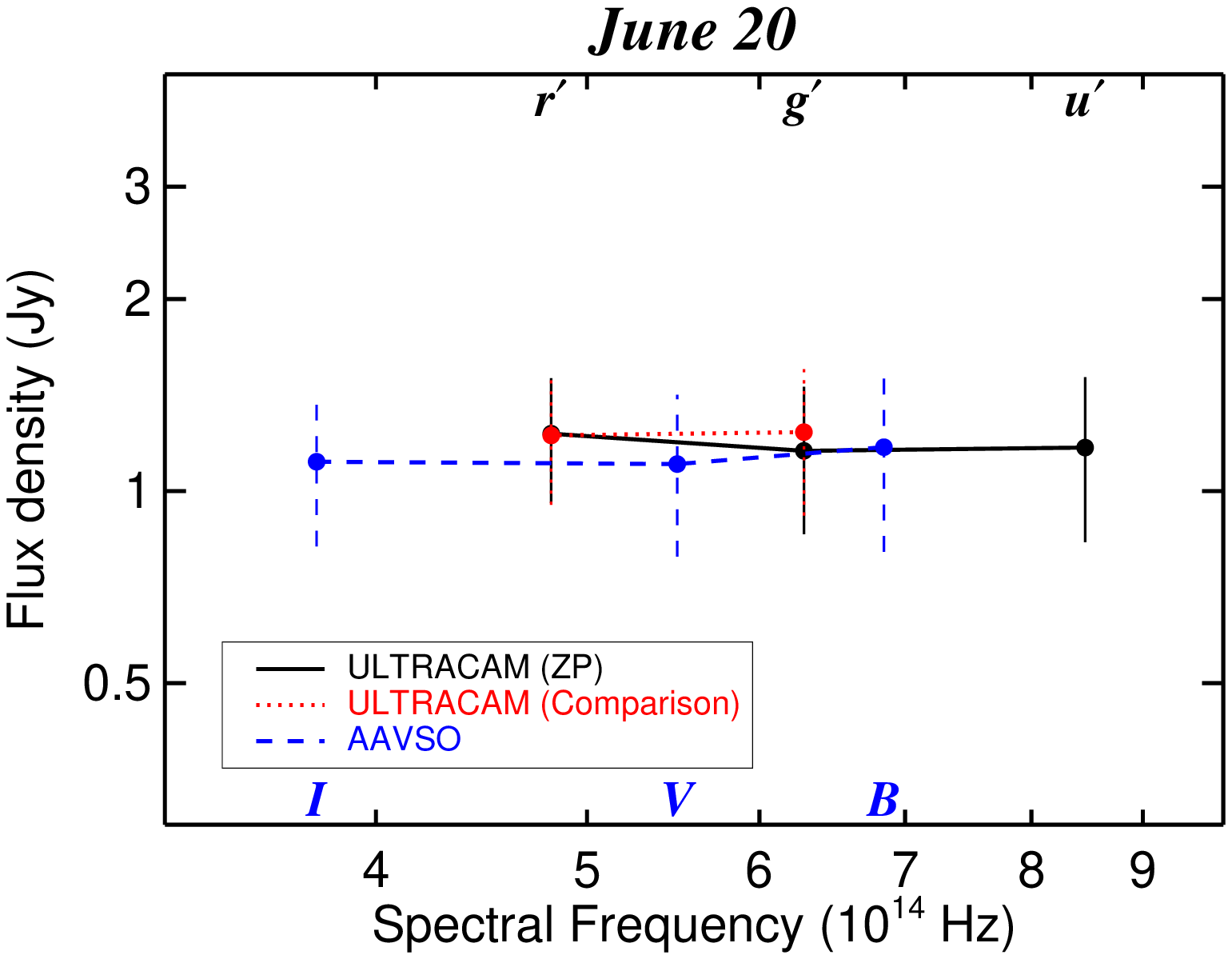}
    \includegraphics[angle=0,width=8.5cm]{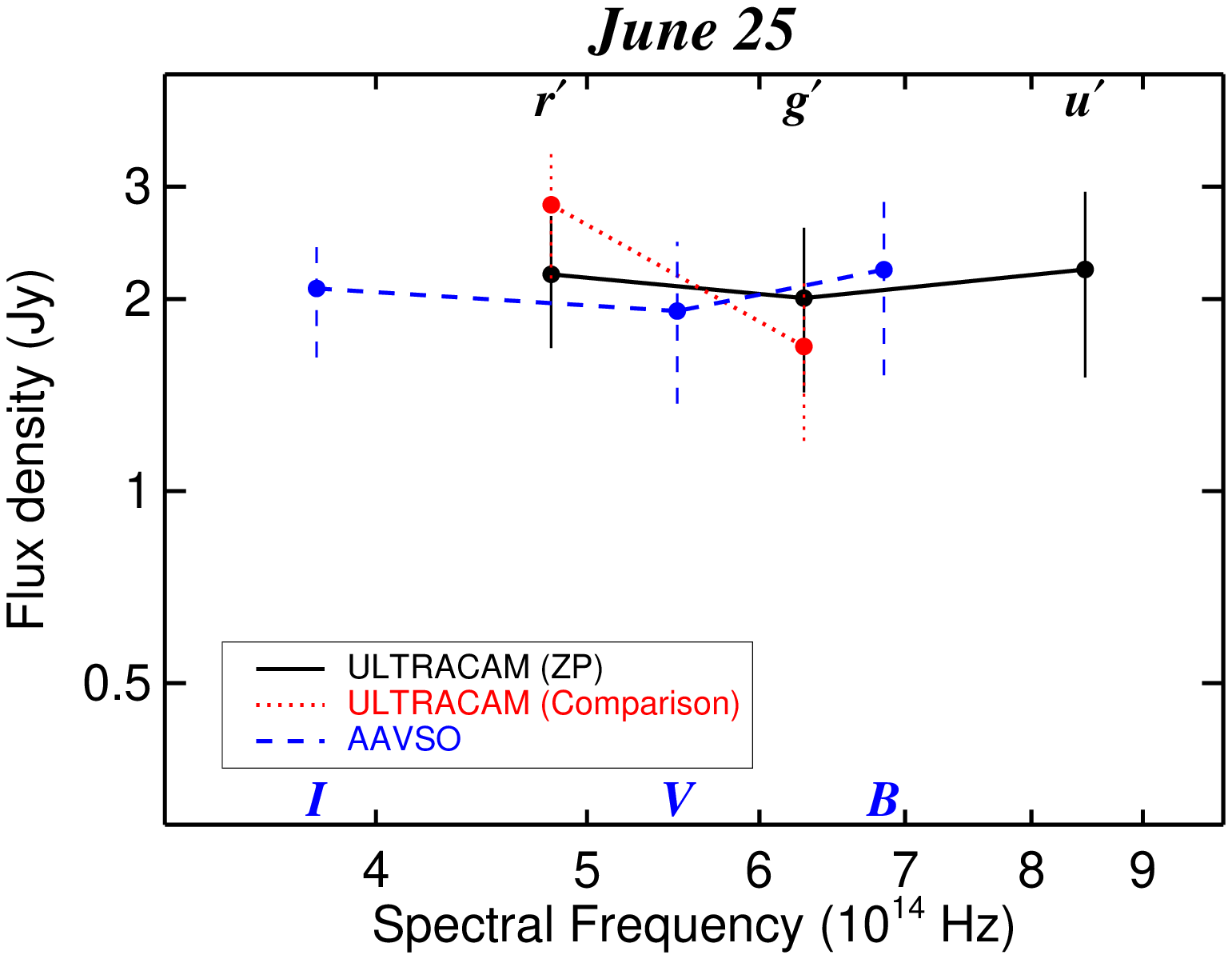}
    \caption{Comparison of three methods to compute SEDs of V404\,Cyg, as a test of the flux calibration used. The nights of June\,20 (Left) and June\,25 (Right) were used because of the availability of useful AAVSO overlap on these nights. The black points and solid curve denote the median SEDs calculated from the standard calibrations based upon zeropoints derived from photometric standards, and used throughout the main body of the paper. The blue points and dashed lines are the SEDs computed from AAVSO data in the Johnsons $B$, $V$ and Cousins $I$ bands. The red dotted points and lines show the ULTRACAM SEDs with the flux calibration derived from the reported magnitudes of the comparison stars (comparison star 1 for June\,20, and comparison star 2 on June\,25). All SEDs are dereddened assuming \av\,=\,4\,mag and the \citet{cardelli89} law. The error bars show the standard deviation of fluxes, so denote the full range of variability. But it should be noted that since the variations in the bands are correlated, the variations of the SED {\em shape} are milder. There is reasonable agreement between the AAVSO and ULTRACAM photometric standards methods, in particular.
 \label{fig:comparesedmethods}}
  \end{center}
\end{figure*}

AAVSO reported calibrated Vega magnitudes of V404\,Cyg during the 2015 outburst, mostly in the $BVI$ filters in the Johnsons-Cousins system. On June\,20 and June\,25, there are 19 and 175 reported $B$ observations, respectively, overlapping with the ULTRACAM observing window, and many more in the $VI$ filters. The observations were carried out by many different observers, as a result of which the photometric quality of the data can vary. Moreover, the observations are not strictly simultaneous, so there is some uncertainty related to variability between observations. However, the median SEDs computed from these data ought to be representative of the source, at least for the purposes of our approximate cross-check against the ULTRACAM calibrations. We computed the SEDs by interpolating the $VI$ band photometry on to the nearest $B$ band times, and applying dereddening corrections assuming an \av\,=\,4\,mag and the \citet{cardelli89} law, as before. The median AAVSO SEDs are shown in Fig.\,\ref{fig:comparesedmethods} for both the nights of overlap, together with the median ULTRACAM SEDs computed using the zeropoint flux calibration described above. The agreement in both overall flux and SED shape is encouraging, especially when one considers the differing bands used, and all the caveats mentioned above regarding the AAVSO comparison.

Finally, we also cross-checked our ULTRACAM flux calibration using the known magnitudes of the comparison stars. \citet{udalski91} list magnitudes of $V$\,=\,12.817 and $B$\,=\,13.523 for comparison star 1. Using the photometric transformation equations provided by \citet{smith02_ugriz}, we find SDSS magnitudes of $g'$\,=\,13.128 and $r'$\,=\,12.626 ($U$ and/or $u'$ magnitudes are generally not available for field stars). The SED of V404\,Cyg using these SDSS comparison star magnitudes for count rate cross-calibration is also shown in Fig.\,\ref{fig:comparesedmethods}, and show good agreement with the other methods.

For comparison star 2, the reported $g'$\,=\,13.372 and $r'$\,=\,12.307 in the APASS system \citep{zacharias15}. However, using our zeropoints to obtain calibrated magnitudes, we estimate $g'$\,=\,12.69\p0.01 and $r'$\,=\,12.41\p0.01. The $g'$ magnitude is brighter than reported by a very large offset of $\approx$\,0.7\,mag. The reason for this discrepancy is unclear, but we note that the APASS-reported $g'$ mag is based upon a single observation with no associated uncertainty. If we instead use the reported $B$\,=\,13.39\p0.40 and $V$\,=\,12.438\p0.11, the transformed $g'$\,=\,12.88\,\p\,0.22 agrees within uncertainty with our prediction of $g'$\,=\,12.69. The transformed $r'$\,=\,12.14\p0.24, on the other hand, is mildly discrepant (brighter) by $\approx$\,0.27 mag. The resultant V404\,Cyg SEDs using these calibrations are also shown in Fig.\,\ref{fig:comparesedmethods}, as the red points and dotted lines, and appears redder than derived from the other methods on June\,25. The steeper SED on June\,25 is entirely ascribable to the difference of $\approx$\,0.27\,mag in the mean values of predicted and transformed $r'$ values. We confirmed that the comparison star magnitudes have inherent disparities by additionally using the IPHAS2 catalogue reported Vega magnitudes of $R$\,=\,12.39, $I$\,=\,11.94\footnote{There are no reported uncertainties specific to these magnitudes, but the reported overall external precision is 0.03\,mag \citep{iphas2}.} and transforming these to SDSS $r'$ using the equations of \citet{jordi06}. We find $r'$\,=\,12.60, which is, indeed discrepant (fainter) by $\approx$\,0.46\,mag than the APASS-transformed mag or $r'$\,=\,12.14 above, and 0.19\,mag fainter than our prediction of $r'$\,=\,12.41. Whether these differences are due to measurement problems or due to the star being variable is unknown. 

These comparisons suggest differences in the flux calibration of $\approx$\,0.2 in $g'$ and $r'$, especially on the nights of June\,21--26. Given the issues with the reported comparison star magnitudes discussed above, the fact that these are not available in $u'$, and the reasonable (approximate) agreement of the ULTRACAM photometric standard based SEDs with the AAVSO SEDs, we prefer the ULTRACAM photometric standard zeropoints for our work. However, this systematic uncertainty should be kept in mind if one is interested in the exact SED shapes. Since these are systematic uncertainties, they do not affect any of the discussion related to the relative {\em variability} in fluxes and spectral indices.

\subsection{Light curves overlay for June\,26} 

\begin{figure*}
  \begin{center}
    \includegraphics[angle=90,width=15cm]{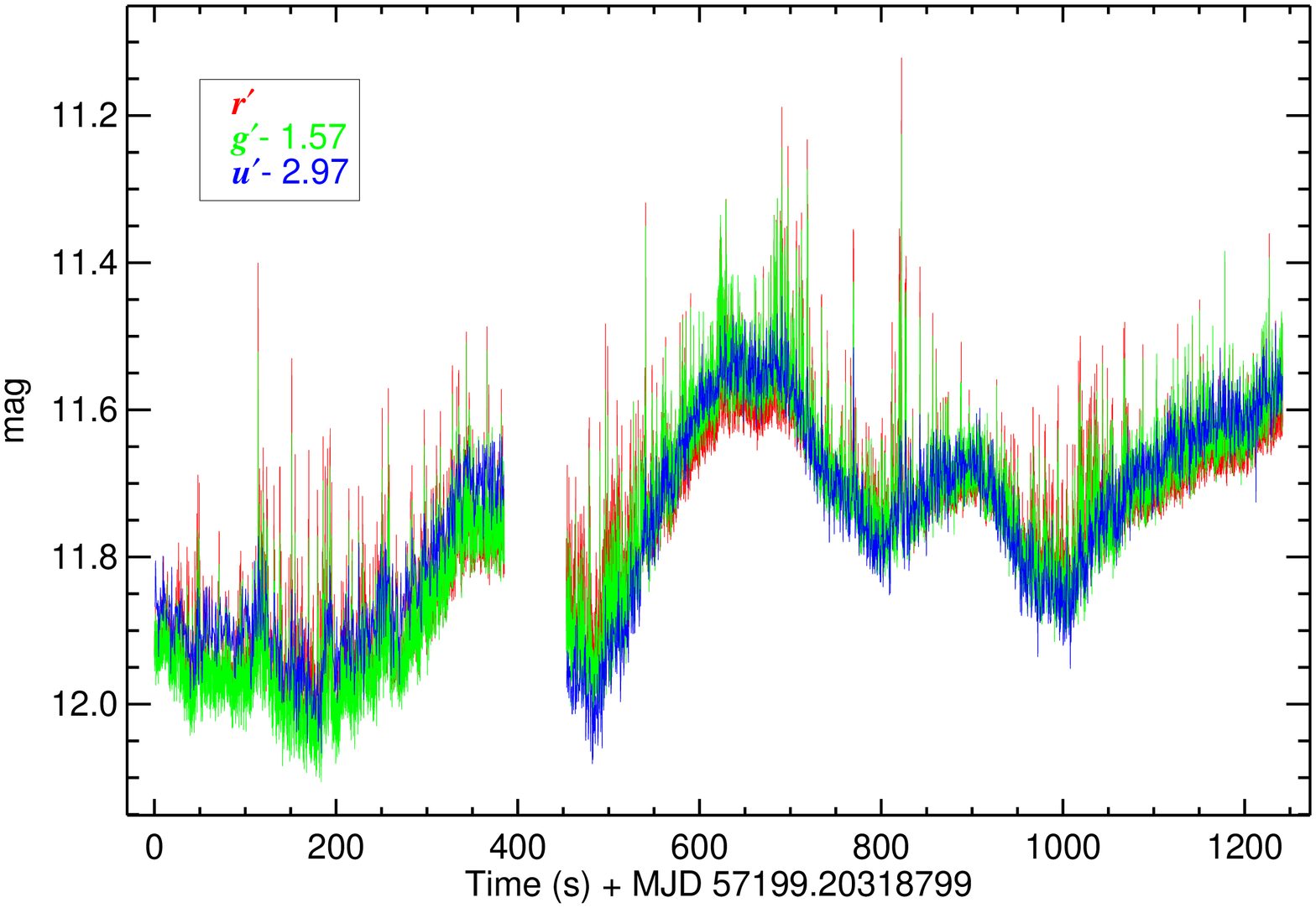}
    \captionsetup[]{singlelinecheck=off}
    \caption{Similar to Fig.\,\ref{fig:lc}, this plot shows the ULTRACAM light curves from 2015 June\,26 (UTC). In this case, the $u'$ and $g'$ light curves have been shifted to the median of the $r'$ lightcurve to aid direct colour comparison. The plot clearly shows the stronger peak-to-peak changes for the slow variations in the bluer bands, contrasted with the stronger $r'$ sub-second flares as compared to $g'$. Note that the $u'$ light curve has lower time resolution by a factor of 15 than the other two bands, so is not suitable for comparing the fast flares with the other bands. 
 \label{fig:lcoverlay}}
  \end{center}
\end{figure*}

Fig.\,\ref{fig:lcoverlay} enlarges the flux-calibrated June\,26 lightcurves to show another perspective. As compared to Fig.\,\ref{fig:lc}, here, the bands have been overlayed on each other fully by shifting $g'$ and $u'$ to the median of the $r'$ light curve for relative comparison. The slow variations are immediately seen to have stronger peak-to-peak variability in the bluer filters. 
We detrended the light curves to isolate the continuum count rate from the fast flares (detailed in section\,\ref{sec:fastflares}), and found maximum peak-to-peak continuum variations of 0.43, 0.42 and 0.35\,mag in $u'$, $g'$ and $r'$, respectively. 
Furthermore, the stronger $r'$ spikes above $g'$ at the times of the fast flares immediately show that the fast flaring is stronger in $r'$ than in $g'$.

\label{lastpage}
\end{document}